\documentclass[letterpaper, 10pt, conference]{IEEEtran}
\IEEEoverridecommandlockouts
\usepackage{amssymb}
\usepackage{cuted}
\usepackage{amsmath,amsfonts}
\usepackage{algorithm}
\usepackage{algpseudocode}
\usepackage{cite}
\usepackage{graphicx}
\usepackage{subcaption}
\usepackage{hyperref}
\usepackage{xurl}
\usepackage{amssymb}
\usepackage{mathrsfs}
\usepackage{array}
\usepackage{mathabx}
\usepackage{xcolor}
\usepackage{mathtools}
\usepackage{linegoal}
\usepackage{caption}
\usepackage{cleveref}
\usepackage{amsthm}

\theoremstyle{definition}

\definecolor{inblue}{HTML}{0072B2}    % same as the figure
\definecolor{outorange}{HTML}{E69F00} % same as the figure

\newtheorem{theorem}{Theorem}
\newtheorem{proposition}{Proposition}
\newtheorem{lemma}{Lemma}
\newtheorem{corollary}{Corollary}
\newtheorem{definition}{Definition}
\newtheorem{problem}{Problem}

\newtheorem{assumption}{Assumption}

\newcommand{\R}{\mathbb{R}}

\newcommand{\C}{\mathbb{C}}

\newcommand{\rank}{\textup{rank}}

\newcommand{\Pcal}{\mathcal{P}}
\newcommand{\Acal}{\mathcal{A}}
\newcommand{\Mcal}{\mathcal{M}}

\newcommand{\Rcal}{\mathcal{R}}
\newcommand{\Bcal}{\mathcal{B}}

\newcommand{\Ccal}{\mathcal{C}}

\newcommand{\Dcal}{\mathcal{D}}

\newcommand{\Scal}{\mathcal{S}}

\newcommand{\tSigma}{\tilde{\Sigma}}

\newcommand{\bV}{\textup{\textbf{V}}}
\newcommand{\bI}{\textup{\textbf{I}}}

\newcommand{\tbf}{\textbf}
\newcommand{\tA}{\tilde{A}}
\newcommand{\ta}{\tilde{a}}
\newcommand{\tmu}{\tilde{\mu}}
\newcommand{\tnu}{\tilde{\nu}}
\newcommand{\tx}{\tilde{x}}

\newcommand{\tB}{\tilde{B}}
\newcommand{\tC}{\tilde{C}}

\newcommand{\tD}{\tilde{D}}

\newcommand{\tM}{\tilde{M}}

\newcommand{\up}{\textup}
\newcommand{\mbf}{\mathbf}

\newcommand{\Z}{\mathbb{Z}}

\newcommand{\LT}{\mathrm{LT}}
\newcommand{\LC}{\mathrm{LC}}

\newcommand{\tquad}{\text{ }}

\newcommand{\tblue}[1]{\textcolor{inblue}{{#1}}}
\newcommand{\torange}[1]{\textcolor{outorange}{{#1}}}

\newcommand{\StateWrap}[1]{
\State \parbox[t]{\dimexpr\linewidth-\algorithmicindent*15/10}{#1}%
}

\DeclarePairedDelimiter{\abs}{\lvert}{\rvert}
\DeclarePairedDelimiter{\abrac}{\langle}{\rangle}

\algtext*{EndIf}
\algtext*{EndFor}
\algtext*{EndWhile}

\title{\LARGE \bf Connectivity-Constrained Linear System Realization: \\ Feasibility and Realizable Sets}

\author{Yaozhi Du, Jing Shuang (Lisa) Li
\thanks{The authors are with the Department of Electrical Engineering and Computer Science, University of Michigan.  
 {\small \{yzdu, jslisali\}@umich.edu}}
}
  
\begin{document}

\maketitle

% Terminology:
% Connectivity pattern (for more "introduction" stuff)
% Support (for more technical stuff)

\begin{abstract}

In this paper, we first study the problem of finding a state-space realization of a given transfer function that adheres to a given connectivity pattern (i.e., support).
While variants of this problem have been studied in distributed and structured control, existing work typically focuses on connectivity patterns relating to some networked plant.
In contrast, we are interested in \textit{arbitrary} connectivity patterns, which are relevant to modeling problems in sensorimotor neuroscience.
We show that this problem is equivalent to solving a system of multivariate polynomial equations, and can be solved exactly without relaxation or approximation.
We then study the reverse problem: given a connectivity pattern, what transfer functions have a state-space realization that adheres to this pattern?
In other words, what transfer functions can this pattern realize?
A surprising finding is that any given pattern can realize nearly all transfer functions, or very few transfer functions.
We provide rigorous results and algorithms for both problems, leveraging technical machinery from algebraic geometry.
The efficacy of our approach is demonstrated in examples at the end of the paper.

\end{abstract}

\section{Introduction}

Feedback controllers are commonly used to model animal sensorimotor behaviors such as reaching \cite{diedrichsen2010coordination, todorov2004optimality}.
How are such controllers implemented in the animal's nervous system?
The prevailing view in motor neuroscience is that the nervous system performs computations underlying control through dynamics \cite{vyas2020computation, li2025toward}.
In control-theoretic terms, the controller is \textit{realized} on neural dynamics.
This is convenient for theorists, as it allows us to leverage standard realization techniques for modeling.
Less conveniently, biological networks are sparsely connected \cite{bullmore2012economy}, imposing connectivity constraints on the realizations and complicating the modeling problem.
While variants of the controller realization problem have been studied in distributed and structured control, existing work typically focuses on connectivity patterns relating to some networked plant; in contrast, in the biological setting, connectivity constraints are fixed by anatomy and likely unrelated to connectivity of the plant (i.e., body).
Thus, in this paper, we study the problem of finding connectivity-constrained realizations for arbitrary connectivity patterns and controllers.

Many approaches exist for obtaining controllers with sparse connectivity patterns.
Some approaches use structured realizations based on Youla or Youla-like parameterizations \cite{lessard2013structured, elia2015distributed, vamsi2015optimal, sabuau2023network}; others use alternative parameterizations such as System Level Synthesis (SLS) \cite{anderson2019system,du2026distributed}.
These approaches typically design and realize the controller at the same time; they cannot be used to realize a given controller.
Lower order sparse systems (either controllers or plants) can be obtained from higher order sparse systems using structured approximation and reduction \cite{sandberg2009model, papachristodoulou2010structured, cheng2019balanced}.
Another approach is sparsity-promoting control \cite{fardad2011sparsity, babazadeh2016sparsity,dhingra2016method}, which penalize density in the controller (i.e., number of communication links in a distributed controller).
This has been applied to power systems\cite{dorfler2014sparsity}\cite{massenio2020data} and spatially invariant systems \cite{zoltowski2014sparsity}. 
However, there is no guarantee that the obtained controllers satisfy a prescribed connectivity pattern. 

The majority of these works are for networked cyber-physical systems. 
Consequently, they typically require that the connectivity pattern of the controller is related (e.g., matched) to the connectivity pattern of some networked plant; this is especially apparent in the SLS parameterization \cite{li2020separating}.
While this requirement is valid for cyber-physical systems, it does not apply for sensorimotor systems, in which the connectivity pattern of the controller is dictated by neural anatomy, not the plant.
Thus, in this work, we remove this requirement entirely and allow the connectivity pattern to be arbitrary.
We first study the problem of whether a given controller can be implemented on a given connectivity pattern (Problem~\ref{pb:complex_realization},~\ref{pb:real_realization}).
Then, motivated by the presence of canonical microcircuits in animal nervous systems \cite{capone2016canonical, li2025toward} (i.e., connectivity patterns that common occur and are thought to perform many different tasks), we study the problem of which controllers a given connectivity pattern can realize (Problem~\ref{pb:generically_realizing}).
We focus on the linear case; linear controllers are widely used in models of neural feedback control \cite{diedrichsen2010coordination, todorov2004optimality}, as are linear models of coarse neural dynamics \cite{nozari2024macroscopic}.

\textit{Contributions:} 
\begin{enumerate}
    \item We provide necessary and sufficient conditions for connectivity-constrained realizability (Theorem \ref{thm:problem_to_poly_eqns}), and associated algorithms to determine whether a transfer function has a state-space realization adhering to a given connectivity pattern (Algorithms \ref{alg:comp_solns} and \ref{alg:real_solns}).
    \item We prove that all connectivity patterns can realize either nearly all transfer functions or very few transfer functions (Corollary \ref{cor:not_gen_realizing}); ``nearly all'' and ``very few'' are made mathematically precise in Section~\ref{sec:generically_realizing}. 
    \item We provide necessary and sufficient conditions to check whether a given connectivity pattern can realize nearly all transfer functions (Theorem \ref{thm:properties_of_generically_realizing}), and the associated algorithm (Algorithm \ref{alg:generically_realizing}).
\end{enumerate}

A preliminary conference version of this work appeared in \cite{du2026sparse}, which focused on contribution 1). 
The present work substantially extends this and adds contributions 2) and 3). 
The paper is organized as follows.
Relevant definitions are introduced in Section~\ref{sec:setup}. Section~\ref{sec:equivalence} shows that solving the sparse realization problem is equivalent to solving a system of multivariate polynomial equations; Section~\ref{sec:grobner} introduces tools to solve this.
Section~\ref{sec:generically_realizing} analyzes what transfer functions a given connectivity pattern can realize.
Section~\ref{sec:simulations} includes simulations demonstrating the efficacy of our approach; concluding remarks are provided in Section~\ref{sec:conclusions}.

\textbf{Notation:} $\delta_{ij}$ is an $n\times n$ matrix with the $(i,j)$-th entry equal to $1$ and zeros elsewhere. $\Bbbk[x] \coloneq \Bbbk[x_1,\cdots,x_k]$ is the set of multivariate polynomials in $x_1,\cdots,x_k$ with coefficients in $\Bbbk$, and $GL_n(\Bbbk)$ is the set of $n\times n$ invertible matrices with entries in $\Bbbk$. 
In this paper, $\Bbbk$ means ``$\R$ or $\C$" (this will be apparent from context).
We use letters such as $A$, $x$, to denote general matrices and vectors; we use calligraphic letters $\Acal$ to denote supports of matrices; we use bold letters $\mathbf{K}$ to denote transfer function matrices.
We use these letters with tildes to indicate values satisfying specific properties (e.g., $x=\tilde{x}$ to denote a solution to $f(x)= 0$).

\section{Problem setup} \label{sec:setup}

Consider linear time-invariant (LTI) controllers of the form:
\begin{equation} \label{equ:abcd_realization}
    \begin{aligned}
        \dot{z}(t) &= Az(t)+By(t)\\
        u(t) &= Cz(t)+Dy(t)
    \end{aligned}
\end{equation}
with sensory input $y(t)$ and control output $u(t)$, where
$A\in\C^{n\times n}$, $B\in\C^{n\times m}$, $C\in\C^{q\times n}$, and $D\in\C^{q\times m}$. 
We denote this system concisely as:
\begin{equation}
    \Sigma = (A,B,C,D) \label{equ:ss_realization_sigma}
\end{equation}

The $m \times q$ transfer function matrix \textbf{realized} by $\Sigma$ is $\mbf{K}$\footnote{A remark on notation: we write $\mbf{K}$ instead of $\mbf{K}(s)$, suppressing Laplace variable $s$ for readability. The argument slot is instead reserved for design parameters: in later sections, we write $\mbf{K}(a)$ to denote the dependence of the transfer function on parameters $a$ (rather than frequency).}: 
\begin{equation} \label{eq:realization_equation}   
    \mbf{K} = C(sI-A)^{-1}B + D
\end{equation}
In this paper we consider transfer function matrices with no pole-zero cancellations.
From standard results, a given transfer function matrix has infinite realizations.
A \tbf{minimal realization} is one with the smallest possible $A$ matrix; in this paper we consider only minimal realizations.

\textbf{Remark:} While we focus our discussion on continuous-time controllers, our approach applies to arbitrary causal linear systems in both continuous- and discrete-time. 

We now provide a more formal definition on connectivity pattern.

\begin{definition} \label{def:support}
    Given a state space system ${\Sigma} = (A,B,C,D)$:
    \begin{enumerate}
        \item \tbf{Supports} compatible to ${\Sigma}$ (in terms of dimension) can be expressed as:
        \begin{equation}
            \begin{aligned}
            &\Scal = (\Acal,\Bcal,\Ccal,\Dcal),\quad \Acal\in\{0,1\}^{n\times n}, \\
            &\Bcal\in\{0,1\}^{n\times m},\quad \Ccal\in\{0,1\}^{q\times n},\quad
            \Dcal\in\{0,1\}^{q\times m}
            \end{aligned}
        \end{equation}  
        \item ${\Sigma}$ \textbf{satisfies} a support ${\Scal} = ({\Acal},{\Bcal},{\Ccal},{\Dcal})$ if
        \begin{equation}
            \begin{aligned}
                &\forall i,j,\text{ }{\Mcal}_{ij} = 0,\text{ } M_{ij} = 0\\
                &\Mcal\in\{{\Acal},\Bcal,\Ccal,\Dcal\},\text{ } M\in\{A,B,C,D\}
            \end{aligned}
        \end{equation}
        \item $\Scal$ is the support of $\Sigma$ if
        \begin{equation}
            \begin{aligned}
                &\forall i,j,\text{ }{\Mcal}_{ij} = \begin{cases}
                    1,\quad M_{ij}\neq 0 \\
                    0,\quad M_{ij} = 0
                \end{cases}\\
                &\Mcal\in\{\Acal,\Bcal,\Ccal,\Dcal\},\text{ } M\in\{A,B,C,D\}
            \end{aligned}
        \end{equation}
    \end{enumerate}
\end{definition}
In this paper, we refer to \textit{connectivity patterns} and \textit{supports} interchangeably.
We now formalize the problems under consideration:

\begin{problem}
    Given a controller ${\mbf{K}}$ and a support $\Scal$, find a realization $\Sigma$ of $\mbf{K}$ that satisfies $\Scal$. \label{pb:complex_realization}
\end{problem}

\begin{problem}
    Given a controller $\mbf{K}$ and a support $\Scal$, find a \textit{real-valued} realization $\Sigma$ of $\mbf{K}$ that satisfies $\Scal$. \label{pb:real_realization}
\end{problem}

Problem~\ref{pb:real_realization} is introduced because many applications (including neuroscience) are concerned with real-valued realizations. 
While Problems~\ref{pb:complex_realization} and~\ref{pb:real_realization} are concerned with specific controller, certain supports can be used to realize \textit{many} different compatible controllers; we are also interested in the following problem:

\begin{problem}
    Given a support $\Scal$, find the set of controllers $\mbf{K}$ with at least one realization $\Sigma$ satisfying $\Scal$. \label{pb:generically_realizing}
\end{problem}

Depending on controller $\mbf{K}$ and support $\Scal$, solutions to Problems \ref{pb:complex_realization} and \ref{pb:real_realization} may or may not exist; this is formalized by the following:

\begin{definition}
   A controller $\mbf{K}$ is (real-)\tbf{realizable} on a support $\Scal$ if there exists (real-valued) $\Sigma$ of $\mbf{K}$ that satisfies $\Scal$.
\end{definition}

\section{Equivalence of sparse realization to multivariate polynomial equations} \label{sec:equivalence}

We now tackle Problems \ref{pb:complex_realization} and \ref{pb:real_realization} and show that the existence of controller realizations is equivalent to the existence of solutions to certain multivariate polynomial equations.

First, recall the canonical result that we can convert between minimal realizations using similarity transforms:

\begin{lemma}[{[\citen{de2000minimal}], Prop 2.3}]
\label{lem:similarity_of_min_realizations}
    Let $\Sigma_1 = (A_1,B_1,C_1,D_1)$ and ${\Sigma_2} = (A_2,B_2,C_2,D_2)$ be minimal realizations of $\mbf{K}$. $\exists T \in {GL}_n(\C)$ s.t.
    \begin{equation}
        A_2 = T^{-1} A_1 T, \text{ } B_2 = T^{-1} B_1, \text{ } C_2 = C_1 T,\text{ } D_2 = D_1
    \end{equation}
    In particular, if $\Sigma_1$ and $\Sigma_2$ are real-valued, $T\in{GL}_n(\R)$.
\end{lemma}

Lemma~\ref{lem:similarity_of_min_realizations} implies that matrix $D$ is the same for all minimal realizations of controllers $\mathbf{K}$. Then:

%Thus, when we find one with $D$ not satisfying the desired support, we know a realization does not exist, formally written as: 

\begin{corollary} \label{corr:D}
    Given a controller $\mbf{K}$ and a support $\Scal$, consider $D = \lim_{s \rightarrow \infty}\mbf{K}$. If there exists $i,j$ such that $\Dcal_{ij} = 0$ and $D_{ij}\neq 0$, $\mbf{K}$ is not realizable on $\Scal$.
\end{corollary}

For the rest of the paper, we assume that:  
\begin{assumption}
    For any controller $\mbf{K}$ to be realized on some given support $\Scal$, the direct current (DC) component $D = \lim_{s \rightarrow \infty}\mbf{K}$ is such that for all $\Dcal_{ij} = 0$, $D_{ij} = 0$.
    \label{ass:dc_component_satisfies}
\end{assumption}

By Corollary~\ref{corr:D}, any controller $\mbf{K}$ is not realizable on $\Scal = (\Acal,\Bcal,\Ccal,\Dcal)$ if it does not satisfy Assumption~\ref{ass:dc_component_satisfies}. 
Also, for all controllers that satisfy assumption~\ref{ass:dc_component_satisfies}, $D$ remains unchanged across all realizations, so we can separate out this DC component and analyze the realization of the remaining dynamics (which will be strictly proper) on $(\Acal,\Bcal,\Ccal,0)$; we will do this for the remainder of the paper.

By Lemma~\ref{lem:similarity_of_min_realizations}, any two minimal realizations are related by a similarity transform, which means:

\begin{corollary}
    \label{thm:problems_equal_finding_T}
    Let $\Sigma = (A,B,C,0)$ be a minimal realization of $\mbf{K}$, let $\Sigma_{T} \coloneq (T^{-1}AT,T^{-1}B,CT,0)$ denote its transformation under some $T\in{GL}_n(\C)$. Then,
    \begin{enumerate}
        \item A realization of $\mbf{K}$ satisfying $\Scal$ exists if and only if for some $\Sigma$, there exists $T\in{GL}_n(\C)$ such that $\Sigma_{T}$ satisfies $\Scal$.
        \item A real-valued realization of $\mbf{K}$ satisfying $\Scal$ exists if and only if for some $\Sigma$, there exists $T\in{GL}_n(\C)$ such that $\Sigma_{T}$ satisfies $\Scal$ and is real-valued. In particular, if $\Sigma$ is real-valued, we require $T\in{GL}_n(\R)$.
    \end{enumerate}
    \label{thm:existence_of_T}
\end{corollary}

In other words, a realization that satisfies $\Scal$ exists if we can find an appropriate transformation of an existing minimal realization $\Sigma$ (which might not have the required support). 
In practice, we can pick $\Sigma$ to be the modal realization or a Jordan form realization.

To find an appropriate transformation (i.e., invertible matrix), we first notice that invertible matrices are products of invertible elementary matrices corresponding to row permutations, row additions, or multiplication of a row by a nonzero scalar; this can be seen through row reduction.
The elementary matrices are:

\begin{enumerate}
    \item Elementary permutation matrix $E^p_{ij}$: identity matrix with rows $i$ and $j$ permuted.
    \item Elementary addition matrix: identity matrix with one off-diagonal entry replaced by nonzero scalar parameter $x$:
    \begin{equation}
        E^a_{ij}(x) = I_n+x\delta_{ij}, \quad E^a_{ij}(x)^{-1} = I_n-x\delta_{ij}, \quad i \neq j \label{equ:elementary_addition}
    \end{equation}
    \item Elementary multiplication matrix $E^m_{i}(x)$: identity matrix with row $i$ multiplied by nonzero scalar $x$.
\end{enumerate}

Without loss of generality, we construct invertible matrices by first applying permutations, then additions, then multiplications. 
Let $T_p$, $T_a$, and $T_m$ denote products of the relevant elementary permutation, addition, and multiplication matrices, respectively. Then:

\begin{equation}
    T_mT_aT_pT = I,\quad T^{-1} = T_mT_aT_p \label{equ:decompose_T}
\end{equation}

Now, consider the process of decomposing some invertible $T$ into $T_p$, $T_a$, and $T_m$. 
In this case, $T_p$ takes finitely many different values. 
We denote the set of possible values as:
\begin{equation}
\begin{aligned}
    \Pcal_p = \{T_p\mid T_p\up{ is the product of elementary}\\ \up{permutation matrices}\}
\end{aligned}
\end{equation}

Then, given $T_pT$ for some $T_p \in \Pcal_p$, we seek $T_m$ and $T_a$. 
These are non-unique.
We construct valid matrices using the following convention: given $T_pT$, process columns from left to right. 
Within each column $i$, cancel nonzero off-diagonal entries from top to bottom by adding appropriate scalar multiples of row $i$.

Using this convention, we present an equivalent statement of realizability of a controller on a support:

\begin{lemma}
    \label{lem:realizability_and_T} 
    Given $\mbf{K}$ with realization $\Sigma=(A,B,C,0)$, $\mbf{K}$ is realizable on support $\Scal$ if and only if there exists a product of elementary permutation matrices $T_p$ and a product of elementary addition matrices $T_a$ s.t. ${\Sigma}_{\Scal}$ satisfies $\Scal$, where
    \begin{equation}
    \begin{aligned}
        &{\Sigma}_{\Scal} = (A_{\Scal},B_{\Scal},C_{\Scal},0),\tquad A_{\Scal} = (T_aT_p)A(T_a T_p)^{-1}\\
        &B_{\Scal} = (T_a T_p)B,\tquad C_{\Scal} = C( T_a T_p)^{-1} \label{equ:tilde_realizations}
    \end{aligned}
\end{equation}\label{lem:drop_Tm}
\end{lemma}

\begin{proof}
    Write $T$ as in \eqref{equ:decompose_T}. Then, using \eqref{equ:tilde_realizations}, we can rewrite Corollary~\ref{thm:problems_equal_finding_T} as:
    $\mbf{K}$ is realizable on $\Scal$ if and only if $\exists$ $T_m$ such that:
    \begin{equation}
        \Sigma_m = (T_m A_{\Scal} T_m^{-1}, T_mB_{\Scal},C_{\Scal} T_m^{-1}, 0)
    \end{equation}
    satisfies $\Scal$. Since $T_m$ is diagonal and invertible, the support of $\Sigma_{\Scal}$ and $\Sigma_m$ are the same.
\end{proof}

Lemma~\ref{lem:drop_Tm} uses the fact that the product of multiplication matrices $T_m$ is irrelevant for sparse state-space realization; thus, for the purpose of finding sparse realizations, we can take $T_m = I$.
Then, the problem of finding a sparse realization translates to finding appropriate $T_p$ and $T_a$.
$T_a$ makes at most $n(n-1)$ row additions. 
By \eqref{equ:elementary_addition}, the entries of elementary addition matrices $E_{ij}^a(x)$ and its inverse $E_{ij}^a(x)^{-1}$ are polynomials of the parameter $x$ with real coefficients. 
We collect these parameters and denote them as:
\begin{equation}
    x\coloneq[x_1, \cdots,x_{n(n-1)}]^\top \in \C^{n(n-1)}
\end{equation}
where $x_i$ is the parameter of the $i-$th elementary addition matrix involved.
This allows us to express the entries of $T_a$ and $T_a^{-1}$ as polynomials of $x$; we will write $T_a(x)$ and $T_a^{-1}(x)$ to explicitly reflect this property.
%\begin{equation}
%\begin{aligned}
%    &T_a\coloneq T_a(x),\quad T_a^{-1} \coloneq T_a^{-1}(x), \\
%    &\left[T_a(x)\right]_{ij}\in\R[x], \quad \left[T_a^{-1}(x)\right]_{ij}\in\R[x]\\
%    &T_a \coloneq T_a(\tx),\quad  T_a^{-1}\coloneq T_a^{-1}(\tx)
%\end{aligned}
%\end{equation}
Then, we rewrite \eqref{equ:tilde_realizations} as:
\begin{equation}
\begin{aligned}
    \Sigma_{\Scal} &= (A_{\Scal}(x), B_{\Scal}(x),C_{\Scal}(x),0) \\
    A_{\Scal}(x) &= T_a(x)T_p A T_p^{-1}T_a^{-1}(x),\\
    B_{\Scal}(x) &= T_a(x)T_p B,\\
    C_{\Scal}(x) &= C T_p^{-1}T_a^{-1}(x)
    \label{equ:ABC_S_x}
\end{aligned}
\end{equation}
where each matrix element (e.g., $\left[A_{\Scal}(x)\right]_{ij}$) is a polynomial in $\C[x]$.
Then, for some given transfer function matrix $\mbf{K}$ with realization $\tSigma = (\tA,\tB,\tC,0)$, given support $\Scal$, and permutation matrix $T_p$, we can collect these polynomials into a set of polynomials:
\begin{equation}
\begin{aligned}
    H_{\Scal,\tSigma,T_p} = \{&\left[\tM_{\Scal}(x)\right]_{ij}\mid \Mcal_{ij} = 0,\\
    &\tM\in\{\tA,\tB,\tC\},\Mcal\in\{\Acal,\Bcal,\Ccal\}\}
    \label{equ:the_equivalent_multvar_polys}
\end{aligned}
\end{equation}
Note that when $\tA$, $\tB$, $\tC$ are real valued, $H_{\Scal,\tSigma,T_p}\subseteq \R[x]$.

We now show that Problems \ref{pb:complex_realization} and \ref{pb:real_realization} can be equivalently stated as solving systems of polynomial equations: 

\begin{theorem}
    \label{thm:problem_to_poly_eqns}
    Given support $\Scal$ and controller $\mbf{K}$ with minimal realization $\tSigma=(\tA,\tB,\tC,0)$,
    \begin{enumerate}
        \item $\mbf{K}$ is realizable on $\Scal$ if and only if there exists $T_p\in\Pcal_p$ and $\tilde{x}\in\C^{n(n-1)}$ such that
        \begin{equation} \label{eq:thm1_eqn}
        \forall h\in H_{\Scal,\tSigma, T_p},\quad h(\tx) = 0
        \end{equation}        
        \item When $\tSigma$ is real-valued, $\mbf{K}$ is real-realizable on $\Scal$ if and only if there exists $T_p\in\Pcal_p$ and $\tilde{x}\in\R^{n(n-1)}$ such that \eqref{eq:thm1_eqn} is satisfied.
    \end{enumerate}
\end{theorem}

\begin{proof}
    For 1), by Lemma~\ref{lem:realizability_and_T}, $\mbf{K}$ is realizable on $\Scal$ if and only if there exists $T_p$ and $T_a$ that converts $\tilde{\Sigma}$ to a realization satisfying $\Scal$.
    According to our previous discussion and \eqref{equ:ABC_S_x}-\eqref{equ:the_equivalent_multvar_polys}, the existence of $T_a$ is equivalent to the existence of a common root for polynomials in the set $H_{\Scal,\tSigma, T_p}$; this gives the desired result.

    For the sufficiency of 2), when $\tA$, $\tB$, $\tC$ are real-valued, $\left[\tA_{\Scal}(x)\right]_{ij}$, $\left[\tB_{\Scal}(x)\right]_{ij}$, $\left[\tC_{\Scal}(x)\right]_{ij}\in \R[x]$. When $\tx\in\R^{n(n-1)}$, clearly $\tA_{\Scal}(\tx)$, $\tB_{\Scal}(\tx)$, $\tC_{\Scal}(\tx)$, and therefore $\tSigma_{\Scal}$ are real-valued. Also, by \eqref{eq:thm1_eqn}, $\tSigma_{\Scal}$ satisfies $\Scal$.
    
    For the necessity of 2), by Corollary~\ref{thm:problems_equal_finding_T}, there exists $T\in {GL}_n(\R)$ that converts $\tSigma$ to a realization $\tSigma_{\Scal}$ satisfying $\Scal$. We can write $T = T_aT_p$. $T_p\in\R^{n\times n}$ implies $T_a\in\R^{n\times n}$. Therefore, $T_a$ can be decomposed into elementary addition matrices with real-valued parameters $\tilde{x} \in\R^{n(n-1)}$ (see discussion before \eqref{equ:ABC_S_x}). Since $\tSigma_{\Scal}$ satisfies $\Scal$, \eqref{eq:thm1_eqn} holds.
\end{proof}

%With Theorem~\ref{thm:problem_to_poly_eqns}, Problems~\ref{pb:complex_realization} and~\ref{pb:real_realization} are reduced to solving multivariate polynomial equations. 

\section{Solving multivariate polynomial equations} \label{sec:grobner}

Solving multivariate polynomial equations (i.e., finding common zeros of a set of polynomials) has been studied in the field of algebraic geometry \cite{cox1997ideals,cox1998using,lasserre2008semidefinite}; we leverage these results to solve Problems~\ref{pb:complex_realization} and~\ref{pb:real_realization}.
Subsections \ref{subsec:complex} and \ref{subsec:real} primarily recapitulate existing results, as most control theorists are likely unfamiliar with these.
Subsection \ref{subsec:sparse} applies these algorithms to the realization problems under consideration.
Throughout the section, $k\in\Z^{+}$ denotes number of variables in the relevant set of equations; for example, $k = n(n-1)$ for $H_{\Scal,\tSigma,T_p}$ in \eqref{equ:the_equivalent_multvar_polys}.
Let $H = \{h_1,\cdots,h_\eta\}$ denote a set of polynomials; we consider $h_i\in\C[x]$ in subsection \ref{subsec:complex} and $h_i\in\R[x]$ in subsection \ref{subsec:real}.

\subsection{Computing complex-valued solutions} \label{subsec:complex}

This section provides relevant background on computing complex-valued solutions to polynomial equations. This culminates in Algorithm \ref{alg:comp_solns}.

\begin{definition} \label{def:degree} Total degree of monomials and polynomials.
\begin{enumerate}
    \item A monomial $x^\alpha \in \Bbbk[x]$ is expressed as:
    \begin{equation}
        x^\alpha \coloneq x_1^{\alpha_1}x_2^{\alpha_2}\cdots x_k^{\alpha_k}, \quad \alpha\in \Z_{\geq 0}^k
    \end{equation}
    The \tbf{total degree} of $x^\alpha$ is:
    \begin{equation}
        \abs{\alpha} = \sum_{i=1}^k\alpha_i \label{equ:abs_alpha}
    \end{equation}
    \item A polynomial $f\in \Bbbk[x]$ is expressed as:
    \begin{equation}
        f = \sum_\alpha c_\alpha x^\alpha,\quad \alpha\in\Z_{\geq 0}^n
    \end{equation}
    The \tbf{total degree} of $f$ is $\max \abs{\alpha}$ such that $c_\alpha\neq 0$.
\end{enumerate}
\end{definition}

\begin{definition}[{[\citen{cox1997ideals}], Chapter 1.4}]
    A set of polynomials $I\subseteq \Bbbk[x]$ is an \tbf{ideal} if it satisfies:
    \begin{enumerate}
        \item $0\in I$.
        \item $\forall f,g\in I$, $f+g\in I$.
        \item $\forall h\in \Bbbk[x]$ and $f\in I$, $fh\in I$.
    \end{enumerate}
\end{definition}

\begin{definition}
    Given $H\coloneq\{h_1,\cdots,h_{\eta}\} \subseteq \Bbbk[x]$, the \tbf{ideal generated by} $h_i$ written as $I = \abrac{h_1,\cdots,h_{\eta}}$ is: 
    \begin{equation}
    \begin{aligned}
        I = \{g\in\Bbbk[x]\mid \exists f_i\in\Bbbk[x] \text{ s.t. }        
        g = \sum_{i=1}^\eta{f_ih_i} \}
    \end{aligned}
    \end{equation}
\end{definition}

\begin{definition}
    A \tbf{monomial order $\succ$ on $\Bbbk[x]$} is a relation on different monomials. For two distinct monomials $x^\alpha$ and $x^\beta$, it determines whether $x^\alpha\succ x^\beta$ or $x^\alpha\prec x^\beta$ and satisfies:
    \begin{enumerate}
        \item $\succ$ is a linear order on $\Z_{\geq 0}^n$.
        \item If $x^\alpha\succ x^\beta$, $x^\beta\succ x^\gamma$, then $x^\alpha\succ x^\gamma$.
        \item Every nonempty subset of $\Z_{\geq 0}^n$ has a smallest element under $\succ$.
    \end{enumerate}
\end{definition}

% When computing Gr\"obner basis, different monomial orders results in different computation time and different Gr\"obner basis.
We now introduce two monomial orders used in this paper. %Other monomial orderings are introduced in \cite[Chapter 2.2]{cox1997ideals}.

\begin{definition}
    With $\alpha,\beta\in\Z_{\geq 0}^k$,
    \begin{enumerate}
        \item Lexicographic (\textbf{lex}) order is defined as: $x^\alpha\succ_{\mathrm{lex}}x^\beta$ if for $\alpha = (\alpha_1,\cdots,\alpha_k)$, $\beta = (\beta_1,\cdots,\beta_k)$ the smallest $j$ such that $\alpha_j\neq\beta_j$ satisfies $\alpha_j>\beta_j$. 
        % \item (grevlex order) $x^\alpha\succ_{\mathrm{grevlex}}x^\beta$ if $\abs{\alpha}>\abs{\beta}$ or $\abs{\alpha}=\abs{\beta}$ and the the largest $j$ where $\alpha_j\neq\beta_j$ satisfies $\alpha_j<\beta_j$.
        \item Graded lexicographic (\textbf{grlex}) order is defined as: $x^\alpha\succ_{\mathrm{grlex}} x^\beta$ if $\abs{\alpha}>\abs{\beta}$ or $\abs{\alpha} = \abs{\beta}$ and $\alpha\succ_{\mathrm{lex}}\beta$.
    \end{enumerate}
\end{definition}

% Computing a Gr\"obner basis with lex order facilitates solving polynomial equations. 
% , whereas computing it with grevlex order usually takes less time
% In the following subsection, we will use grlex to represent polynomials as vectors.

\begin{definition}
    \label{def:multdeg_and_LT}
    Given a polynomial $f = \sum_{\alpha\in U}c_\alpha x^\alpha\in\Bbbk[x]$, where $U = \{\alpha\in\Z_{\geq 0}^k\mid c_\alpha\neq 0\}$, and monomial order $\succ$:
    \begin{enumerate}
        \item The \tbf{leading term} of $f$, denoted by $\LT(f)$, is $c_\beta x^\beta$ such that $\forall \alpha\in U\backslash \{\beta\}$, $x^\beta\succ x^\alpha$. 
        \item When $\LT(f) = c_\beta x^\beta$, the \tbf{multidegree} of $f$, written as $\up{multideg}(f)$, is $\beta\in\Z_{\geq 0}^k$.
        \item When $\LT(f) = c_\beta x^\beta$, the \tbf{leading coefficient} of $f$, written as $\LC(f)$, is $c_\beta$.
    \end{enumerate}
\end{definition}

\begin{definition} \textbf{Leading terms of an ideal}. Fix a monomial order $\succ$ on $\Bbbk[x]$. For an ideal $I\subseteq \Bbbk[x]$, we define $\LT(I)$ as the set of leading terms that belong to nonzero elements of $I$:
    \begin{equation}
            \LT(I) = \{cx^\alpha\mid \exists f\in I\backslash\{0\}\up{ s.t. }\LT(f) = cx^\alpha\}
    \end{equation}
\end{definition}

%$\abrac{\LT(I)}$ is the ideal generated by all possible elements of $\LT(I)$

We are now ready to define the Gr\"obner basis, which is frequently used in the remainder of the paper. 

\begin{definition}
    A finite set $G = \{g_1,\cdots,g_\zeta\}\subseteq \Bbbk[x]$ is a \tbf{Gr\"obner basis} of ideal $I$ if:
        \begin{equation}
            \abrac{\LT(g_1),\cdots,\LT(g_\zeta)} = \abrac{\LT(I)}
        \end{equation}
    In particular, $\emptyset$ is the Gr\"obner basis of the zero ideal $\{0\}$. By \cite{cox1997ideals}, every ideal has a Gr\"obner basis.
\end{definition}

\begin{definition}
    A \tbf{reduced Gr\"obner basis} of an ideal $I$ is a Gr\"obner basis $G = \{g_1,\cdots,g_\zeta\}$ such that for $i = 1,\cdots,\zeta$:
    \begin{enumerate}
        \item $\LC(g_i) = 1$.
        \item No monomial of $g_i$ lies in $\abrac{\LT(G\backslash g_i)}$.
    \end{enumerate}
\end{definition}

We now enumerate some useful properties of Gr\"obner bases:
\begin{proposition}[{[\citen{cox1997ideals}]}]
    For an ideal $I$:
    \begin{enumerate}
        \item Gr\"obner bases $G$ of $I$ depends on the applied monomial order $\succ$.
        \item When $G$ is a Gr\"obner basis of $I$ with respect to a monomial order $\succ$, $\abrac{g_1,\cdots,g_\zeta} = I$.
        \item The reduced Gr\"obner basis of $I$ with respect to a fixed monomial order $\succ$ is unique.  
    \end{enumerate}
\end{proposition}

The study of solutions to polynomial equations is the study of \textit{varieties}, which we are now ready to define. 

\begin{definition}
    \label{def:varieties} Varieties and ideals.
    \begin{enumerate}
        \item The \textbf{affine variety} defined by a set of polynomials $H\subseteq \Bbbk[x]$ is:
        \begin{equation}
            \bV(H) = \{x\in\Bbbk^k\mid h(x) = 0\up{ for all }h\in H\}
        \end{equation}
        \item For an ideal $I\neq \{0\}$, we call $\bV(I)$ a \textbf{proper  variety}.
        \item Given affine variety $V\subseteq \Bbbk^n$, the \textbf{ideal of} $V$ is:
        \begin{equation}
        \begin{aligned}
            \up{\tbf{I}}(V) =\{f\in\Bbbk[x]\mid f(x) = 0\up{ for all } x\in V&\}
        \end{aligned}
        \end{equation}
    \end{enumerate} 
\end{definition}

We now discuss how varieties are related to the previously introduced entities (ideals, Gr\"obner bases).
% \begin{proposition}
%     If $I = \abrac{h_1,\cdots,h_\eta}$, then $\bV(I) = \bV(h_1,\\\cdots,h_\eta)$.
%     \label{prop:two_forms_of_V}
% \end{proposition}
\begin{proposition}[{[\citen{cox1997ideals}]}]
    For $H = \{h_1,\ldots,h_\eta\}$, $G = \{g_1,\ldots,g_\zeta\}$ where $\abrac{h_1,\cdots,h_\eta} = \abrac{g_1,\cdots,g_\zeta} = I$, then $\bV(H) = \bV(G) = \bV(I)$.
    \label{prop:two_forms_of_V}
\end{proposition}

In other words, if two sets generate the same ideal, then they have the same variety.
Specifically, if $\{g_1,\cdots,g_\zeta\}$ is a Gr\"obner basis of $\abrac{h_1,\cdots,h_\eta}$, then solving $h_i=0$, $i = 1,\cdots,\eta$ is equivalent to solving $g_j = 0$, $j = 1,\cdots,\zeta$.
This can be used to determine existence of solutions for $\C[x]$.

\begin{proposition}[{[\citen{cox1997ideals}], Chapter 4.1}]
    \label{prop:weak_nullstellensatz}
    For an ideal $I\subseteq\C[x]$, let $G= \{g_1,\cdots,g_\zeta\}$ be the reduced Gr\"obner basis of $I$ with respect to some monomial order. Then:
    \begin{equation}
        \bV(I) = \emptyset\text{ if and only if }G=\{1\}
        \label{equ:existence_of_complex_solns}
    \end{equation}
\end{proposition}

Propositions~\ref{prop:two_forms_of_V} and~\ref{prop:weak_nullstellensatz} provide an easy test for the existence of solutions to multivariate polynomial equations: given $H = \{h_1,\cdots,h_\eta\}$, compute $G$, the reduced Gr\"obner basis of $I = \abrac{h_1,\cdots,h_\eta}$, and check whether $G=\{1\}$ holds. 

When solutions exist, we can find them by applying the following result on a lex-ordered Gr\"obner basis of $I$:
\begin{proposition}{\up{([\citen{cox1997ideals}], \tbf{The Elimination Theorem})}}
    \label{thm:elimination_thm}
    Let $G = \{g_1,\cdots,g_\zeta\}$ be a Gr\"obner basis of an ideal $I$ with respect to lex order. Define    
    \begin{equation}
        G^k \coloneq G\cap \Bbbk[x_{k}]  \label{equ:Gll}
    \end{equation}
    If there exists $\tilde{x}= \left[\tx_1,\cdots,\tx_{k}\right] \in \bV(I)$, then $\tx_k\in\bV(G^k)$.
\end{proposition}

% More efficient monomial orders can be applied in Proposition~\ref{thm:elimination_thm} to improve computational speed.

Here, $\tx_k$ is a common zero of polynomials $G^k$, and is a partial solution for common zeros of $G$.
Since $G^k$ is a set of univariate polynomials, $\tx_k$ can be found through conventional methods (e.g., root finding).
To find a full solution for $G$, we need to extend the partial solution.
This is done by plugging in $\tx_k$ to $G$, then repeating the process (find Gr\"obner basis and re-apply Proposition \ref{thm:elimination_thm}) for the resulting polynomial set.
This process is summarized in Algorithm~\ref{alg:comp_solns}, which computes partial solutions in descending order.
Since Proposition \ref{thm:elimination_thm}) is necessary but not sufficient, it is possible that a given solution $\xi_k \in\bV(G^k)$ does not extend to a full solution.
The algorithm accounts for this by selecting different candidates $\xi_k$ when necessary.
Here, $\up{ComputeReducedGBasis()}$ means computing the reduced Gr\"obner basis (with respect to lex order) of the ideal $\abrac{h_1,\cdots,h_\eta}$, where $H = \{h_1,\cdots,h_\eta\}$.
This can be done in MATLAB using the command \texttt{gbasis}.

%; however, by the Extension Theorem in \cite[Chapter 3.1]{cox1997ideals}, almost all $\tilde{\xi}_k$ extends to a solution $\tx$. 

\begin{algorithm}[h]
\textbf{Input:} $H = \{h_1,\ldots,h_\eta\}$ \\
\textbf{Output:} Solution $\tx = (\tx_1,\ldots,\tx_k)\in\C^k$ or \tbf{Infeasible}
\caption{Computing solutions for a system of multivariate polynomials}
\label{alg:comp_solns}
\begin{algorithmic}[1]
    \State $G = \up{ComputeReducedGBasis}(H,lex)$
    \For{$j = k$ to $1$}
        \State \textbf{if } $G = \{1\}$ \textbf{then return Infeasible}
        % That left arrow: \gets
        \State $G^j = G\cap\C[x_j]$
        \State \textbf{if } $G^j = \emptyset$ \textbf{then } Randomly pick ${\xi}_j\in\C$
        \State \textbf{else } Find ${\xi}_j\in\bV(G^j)$ 
        % that solves equations generated by $G^j$
        \StateWrap{$\hat{G} = \left.G\right|_{x_j = {\xi}_j}$} 
        \StateWrap{$\bar{G} = \up{ComputeReducedGBasis}(\hat{G},lex)$}
        \State \textbf{if} $\bar{G} = \{1\}$ \textbf{ then} go to line 5
        \State $\tx_j = \xi_j$, $G = \bar{G}$
    \EndFor
    \State \Return $\tx$
\end{algorithmic}
\end{algorithm}
% \vspace{-8pt}

\subsection{Computing real-valued solutions} \label{subsec:real}

This section computes real-valued solutions to polynomial equations. 
Existing methods based on semi-definite programming (SDP) \cite{lasserre2008semidefinite} were designed for polynomial equations with finite solutions; we extend these to make it applicable to equations with infinite solutions, culminating in Algorithm \ref{alg:real_solns}.
Recall that we want to solve for:
\begin{equation}
    h_i = 0,\text{ } i = 1,\ldots,\eta,\text{ } h_i\in\R[x]  \label{equ:polynomial_eqns_real}
\end{equation}

% Note that $h_i\in\R[x]$ can have both real and complex valued solutions. To distinguish between the two, we apply the following notations in the subsection:
% \begin{equation}
%     \begin{aligned}
%         \bV(I) = \{\tx\in \C^k\mid, f(\tx) = 0\up{ for all }f\in I\} \\
%         \bV_\R(I) = \{\tx\in \R^k\mid, f(\tx) = 0\up{ for all }f\in I\}
%     \end{aligned}
% \end{equation}

% Then, $\bV(I)$ and $\bV_\R(I)$ denote the set of complex and real-valued solutions and $\bV_\R(I) = \bV(I)\bigcap\R^k$.

Now, we introduce some relevant definitions.
\begin{definition}
    The \tbf{vector form of polynomial $f$} is the infinite row vector $F$ of its coefficients with respect to grlex order, with $x_k\succ x_{k-1}\succ\cdots \succ x_1$. Let $X$ be the infinite column vector of all monomials arranged in increasing order. Then:
    \begin{equation}
        f = FX
        \label{equ:vector_representation_of_polynomials}
    \end{equation}
\end{definition}

For example, consider $f= x_1^2+x_1x_2-4$; $f = FX$ where
\begin{equation}
\begin{aligned}
    F &= \begin{bmatrix}
        -4 & 0 & 0 & 1 & 1 & 0 & \cdots
    \end{bmatrix} \\
    X &= \begin{bmatrix}
        1 & x_1 & x_2 & x_1^2 & x_1x_2 & x_2^2 & \cdots
    \end{bmatrix}^\top
\end{aligned}
\label{equ:F_and_X}
\end{equation}

Each entry of $F$ corresponds to exactly one monomial, so we can index $F$ by the multidegree (see Definition~\ref{def:multdeg_and_LT}) of the corresponding monomials, i.e., $F \coloneq \left[f_\alpha\right]$, $\alpha\in\Z_{\geq 0}^k$.
The entries of $F$ form an infinite sequence $\{f_\alpha\}_{\alpha\in\Z_{\geq 0}^k}$. For a polynomial $f$ of total degree $d$, we can truncate $F$ and $X$ to the monomials of total degree at most $d$ without losing any information. For example, in \eqref{equ:F_and_X}, $f = \sum_\alpha f_\alpha x^\alpha = \sum_{\abs{\alpha}\leq 2}f_\alpha x^\alpha$.

\begin{definition}
    A \tbf{moment matrix} $M(\gamma)$ of sequence $\{\gamma_\alpha\}_{\alpha\in\Z_{\geq 0}^k}$ is the infinite matrix defined by:
    \begin{equation}
        \left[M(\gamma)\right]_{\alpha,\beta} \coloneqq \gamma_{\alpha+\beta},\quad \alpha,\beta\in\Z_{\geq 0}^k
    \end{equation}
\end{definition}

Recalling the definition of total degree $\abs{\alpha}$ from Definition \ref{def:degree}, we can now define truncated moment matrices.

\begin{definition}
    A \tbf{truncated moment matrix} $M_t(\gamma)$ of $M(\gamma)$ is a finite submatrix of $M(\gamma)$ that preserves $\left[M(\gamma)\right]_{\alpha,\beta}$ for $\abs{\alpha}\leq t$ and $\abs{\beta} \leq t$.
\end{definition}

\begin{definition}
    Given a polynomial $f = \sum_{\beta}f_{\beta}x^\beta$ and a sequence $\{\gamma_\alpha\}$, the sequence $\{(f\gamma)_\alpha\}_{\alpha\in\Z_{\geq 0}^k}$ is defined by:
    \begin{equation}
        (f\gamma)_\alpha = \sum_{\beta}f_\beta \gamma_{\alpha+\beta},\quad \alpha\in\Z_{\geq 0}^k
    \end{equation}
\end{definition}

With these, we can now describe the SDP-based approach.

\begin{proposition}[{[\citen{lasserre2008semidefinite}]}]
    \label{thm:real_solns_criterion}
    Given $H = \{h_1,\cdots,h_\eta\}$, consider the following SDP
    % Solving for 
    \begin{equation}
    \begin{aligned}
        \mathrm{argmin}_\gamma \quad 1 \quad s.t.\quad& \gamma_0 = 1,\quad M_t(\gamma)\succeq 0, \\ &M_{t-d_j}(h_j\gamma)=0,\quad t\geq d\label{equ:feas_sdp}
    \end{aligned}
    \end{equation}
    where
    \begin{equation}
        d_j \coloneq \lceil \up{multideg}(h_j)/2\rceil,\text{ } d\coloneq{\max}(d_j),\text{ } j = 1,\cdots,\eta \label{equ:def_of_d}
    \end{equation}
    % We have:
    \begin{enumerate}
        \item If \eqref{equ:feas_sdp} is infeasible for $t = t\geq d$, then no solutions to \eqref{equ:polynomial_eqns_real} exist.
        \item If \eqref{equ:feas_sdp} if feasible for $t$, we can build $M_{t}({\gamma})$. Then, for $1\leq s\leq t$, let $M_{s}({\gamma})$ be truncations of $M_{t}({\gamma})$. If
        \begin{equation}
            \exists 1\leq s\leq t\quad{s.t.}\quad\rank M_{s}({\gamma}) = \rank M_{s-1}({\gamma}) \label{equ:rank_condition}
        \end{equation}
        we can compute solution candidates for \eqref{equ:polynomial_eqns_real} (detailed below). If \eqref{equ:polynomial_eqns_real} has finite solutions, there exist $s, t$ such that the solution candidate solves \eqref{equ:polynomial_eqns_real}.
    \end{enumerate}
\end{proposition}

The method for computing solution candidates for a given $\Scal$ is presented directly in \cite{lasserre2008semidefinite} and is quite notationally heavy, so we summarize the general steps here.
\begin{enumerate}
    \item Locate a set of linearly independent columns of $M_{s}({\gamma})$. 
    \item Extract submatrices of $M_{s}(\gamma)$ based on indices of linearly independent columns.
    \item Invert one of these submatrices and perform matrix multiplication; solution candidates of $x_i$ are eigenvalues of the resulting matrix products.
\end{enumerate}

Algorithm~\ref{alg:real_solns_base} returns the solutions to a system of polynomial equations generated by $H$ when the solution set is finite. By Proposition~\ref{thm:real_solns_criterion}, we can find the solutions by iteratively increasing the value of $t$. Iteration terminates when the number of solutions is zero or finite. To handle the case with infinite solutions, we include a stopping parameter $r$, and terminate iterations when $t = d + r$. 
A larger $r$ makes it more likely that terminating at $t=d+r$ indeed implies we have infinite solutions, but it also increases the running time.

\begin{algorithm}[h]
    \textbf{Input:} $H \coloneq \{h_1,\cdots,h_\eta\}$, $r$  \\
    \textbf{Output:} Solution $\tx = (\tx_1,\cdots,\tx_k)\in\R^k$ or \tbf{Infeasible} or \tbf{InfSolnsLikely}\\
    \caption{Computing real-valued solutions for a system of multivariate polynomial equations with finite solutions}
    \label{alg:real_solns_base}
    \begin{algorithmic}[1]
    \For{$t = d$ to $d+r$}
    \StateWrap{Solve \eqref{equ:feas_sdp}.}
    \State \textbf{if } \eqref{equ:feas_sdp} is infeasible \textbf{then return Infeasible}
    \If{ \eqref{equ:rank_condition} holds}
    \State Extract solution candidates.
    \If{There exists ${\xi}$ such that $h_i(\xi) =  0$}
    \State $\tx \coloneq {\xi}$.
    \State \Return $\tx$
    \EndIf
    \EndIf
    \EndFor
    \State \Return \tbf{InfSolnsLikely}
    \end{algorithmic}
\end{algorithm}

Algorithm~\ref{alg:real_solns} handles the case with infinite solutions using a wrapper around Algorithm~\ref{alg:real_solns_base}. 
Note that when the number of solutions is infinite, there exists some $l\in\{1,\cdots,k\}$ such that we can pick infinite different $\tx_l\in \R$ for $x_l$ to solve \eqref{equ:polynomial_eqns_real}.
Then, we find $l$ and fix the value of $x_l$, and repeat if necessary.
The solution set of the remaining variables becomes finite.
We find $l$ by finding variables that can be freely assigned.
Note that if a variable can be freely assigned in the space of real values, it can also be freely assigned in the space of complex values.
Then, by Proposition~\ref{thm:elimination_thm}, $l$ is a candidate if $G^l = G\cap\R[x_l] = \emptyset$; we test this candidate by assigning a value to $\tx_l$ and running Algorithm~\ref{alg:real_solns_base}.
If Algorithm~\ref{alg:real_solns_base} returns \textbf{InfSolnsLikely}, we repeat the process to find other values of $l$.
If Algorithm~\ref{alg:real_solns_base} returns \tbf{Infeasible}, this candidate is invalid.
Otherwise, we have found a solution.
On Line 6 of this algorithm, $lex_l$ refers to a lex order where $x_j\succ x_l$ for $j\neq l$ instead of $x_1\succ x_2\succ\ldots\succ x_k$.

% In Algorithm~\ref{alg:real_solns}, a more efficient but less rigorous approach to validate if $\tilde{\xi}_l$ extends to a solution $\Xi\in\R^k$ is to blend in some computations to compute complex valued solutions in Algorithm~\ref{alg:comp_solns} by computing the Gr\"obner basis of $\left.\hat{G}\right|_{x_l=\tilde{\xi}_l}$. We repeat this process until we get polynomial equations with finite complex solutions. The issue with this alternative approach is that polynomials with nonzero finite complex-valued solutions could have no real-valued solution. 

\begin{algorithm}[h]
    \caption{Computing real-valued solutions for a system of multivariate polynomial equations}
    \textbf{Input:} $H \coloneq \{h_1,\cdots,h_\eta\}$, $r$ \\
    \textbf{Output:} Solution $\tx = (\tx_1,\cdots,\tx_k)\in\R^k$ or \tbf{Infeasible}\\
    %\vspace{-12pt}
    \begin{algorithmic}[1]
    \State Compute reduced Gr\"obner basis $G$ of $H$
    % \State \textbf{if } $G = \{1\}$ \textbf{then return Infeasible}
    \State a2Out = Algorithm~\ref{alg:real_solns_base}($G$, $r$)
    \If{a2Out != \textbf{InfSolnsLikely}}
    \State \Return \tbf{a2Out}
    \EndIf
    \For{$l=1$ to $k$}
    \State $\bar{G} = \up{ComputeReducedGBasis}(G,{lex_l})$
    \State $\bar{G}^l = \bar{G}\cap \Bbbk[x_l]$
    \If{$\bar{G}^l = \emptyset$}
    \State Randomly pick $\xi_l$.
    \State $\hat{G} = \left.\bar{G}\right|_{x_l = \xi_l}$.
    \State a2Out = Algorithm\ref{alg:real_solns_base}($\hat{G}$, $r$).
    \If{a2Out = \tbf{Infeasible}}
    \State Go to line 9.
    \EndIf
    \If{a2Out returns a solution}
    \State Combine solution with $\xi$ to form $\tx$
    \State \Return{$\tx$}
    \EndIf
    \EndIf
    \EndFor
    \end{algorithmic} 
    \label{alg:real_solns}
\end{algorithm}

\textbf{Remark:} From numerical experiments in Section \ref{sec:simulations}, we observed that we could often extract real-valued solutions using Algorithm~\ref{alg:comp_solns} despite it not being guaranteed to return real-valued solutions.

\subsection{Finding sparse realizations} \label{subsec:sparse}

We now present Algorithm~\ref{alg:find_realizations}, which solves Problems \ref{pb:complex_realization} and \ref{pb:real_realization}: given a controller $\mbf{K}$ and a support $\Scal$, find a realization (real-valued, in the case of Problem \ref{pb:real_realization}) of $\mbf{K}$ that satisfies $\Scal$.
This algorithm leverages the previously presented algorithms; set input variable \tbf{complex} to \textbf{true} to solve Problem~\ref{pb:complex_realization}, and to \textbf{false} to solve Problem~\ref{pb:real_realization}.
Note that this algorithm nominally iterates over all members $T_p \in \mathcal{P}_p$; however, in practice, setting $T_p = I$ often suffices (this is the case for all conducted numerical experiments in Section \ref{sec:simulations}).
% For larger systems, the approach listed in this section can be time-consuming. In the next section, we will provide a more efficient approach for certain supports.

\begin{algorithm}[h]
    \textbf{Input:} $\mbf{K}$, $\Scal = (\Acal,\Bcal,\Ccal,\Dcal)$, Boolean variable \tbf{complex}, $r$ \\
    \textbf{Output:} A realization $\tSigma_{\Scal} = (A_{\Scal},B_{\Scal},C_{\Scal},D_{\Scal})$ of $\mbf{K}$ satisfying $\Scal$ (real-valued if \tbf{complex} $=$ false) or \tbf{Infeasible}\\
    %\vspace{-12pt}
    \caption{Computing sparse realizations for a given transfer function matrix}
    \label{alg:find_realizations}
    \begin{algorithmic}[1]
    \State Construct a realization $\tSigma = (\tA,\tB,\tC,\tD)$ of $\mbf{K}$.
    \State \textbf{if } $\tD$ does not satisfy $\Scal$ \textbf{then return Infeasible}
    \For{$T_p\in\Pcal_p$}
    \State Construct $H_{\Scal,\tSigma,T_p}$. 
    \If{\tbf{complex} = \tbf{true}}
    \State algOut = Algorithm~\ref{alg:comp_solns}($H_{\Scal,\tSigma,T_p}$)
    \Else
    \State algOut = Algorithm~\ref{alg:real_solns}($H_{\Scal,\tSigma,T_p}$, $r$)
    \EndIf
    \If{algOut contains a solution $\tx$}
    \State \Return $\tilde{\Sigma}_{\Scal}=(A_{\Scal}(\tx),B_{\Scal}(\tx),C_{\Scal}(\tx),\tD)$.
    \EndIf
    \EndFor
    \State \textbf{return Infeasible}
    \end{algorithmic}
\end{algorithm}

\section{Realization capability of a given support}
% Realizable Controllers for A Given Support
\label{sec:generically_realizing}

In this section, we focus on Problem~\ref{pb:generically_realizing}, which is finding the set of controllers that are realizable on a given support. As in Section~\ref{sec:equivalence}, we only consider strictly proper controllers. We show that the set of controllers a support realizes is either ``almost all" of the parameter space (to be made mathematically precise) or is contained in a lower-dimensional subset (e.g., hyperplane). We also develop tools to determine which of these two cases applies for a given support. Our results are summarized by Algorithm~\ref{alg:generically_realizing}  at the end of the section.

\subsection{Generic subsets of an affine space}

In this subsection, we work with $x \in \Bbbk^k$ (i.e., $\mathbb{R}^k$ or $\mathbb{C}^k$).
We now formalize our notion of subsets that cover ``almost all" of $\Bbbk^k$.

\begin{definition}
    Set $W \subseteq \Bbbk^k$ is a \tbf{generic subset} of $\Bbbk^k$ if there exists a nonzero ideal $I\subseteq \Bbbk[x]$, such that:
    \begin{equation}
         \Bbbk^k\backslash\bV(I) \subseteq W
    \end{equation}
\end{definition}

%We use the term generic subset because ``generic" was used to describe sets with identical characters in \cite{shields2003structural}. 
Recall from Definition~\ref{def:varieties} that variety $\bV(I)$ is the set of common zeros to polynomials in $I$. When $I\neq\{0\}$, $\bV(I)$ is contained on a hypersurface.
Then, $\Bbbk^k\backslash\bV(I)$ is the entire space excluding this variety; this is nearly the entire space.

We next show that a generic subset cannot be contained in a proper variety. 
This property will be later used to determine the realization capability of a given support.
\begin{lemma}
    \label{lem:Zariski_Closure_of_Generic_Subsets}
    Let $W$ be a generic subset of $\Bbbk^k$. Then, for all nonzero ideals $I \subseteq \Bbbk[x]$ and varieties $\bV(I)$, $W\nsubseteq\bV(I)$.
\end{lemma}

We prove Lemma \ref{lem:Zariski_Closure_of_Generic_Subsets} by showing that the only variety containing generic subsets of $\Bbbk^k$ is $\Bbbk^k$ itself.
To do so, we must first introduce some more definitions and results.

\begin{definition}
    For $W\subseteq \Bbbk^n$, the \tbf{Zariski Closure} of $W$, denoted $\overline{W}$, is the smallest affine variety in $\Bbbk^n$ containing $W$.
\end{definition}
\begin{definition}
    An ideal $I\subseteq \Bbbk[x]$ is \tbf{prime} if $\forall f,g\in \Bbbk[x]$ and $fg\in I$, either $f\in I$ or $g\in I$.
\end{definition}
\begin{definition}
    An affine variety $V\subseteq \Bbbk^n$ is \tbf{irreducible} if for all affine varieties $V_1$, $V_2$ such that $V = V_1\cup V_2$, we have  $V = V_1$ or $V=V_2$.
\end{definition}
Prime ideals and irreducible varieties are related by:
\begin{proposition}[{[\citen{cox1997ideals}, Chapter~4]}] \label{prop:irreducible_prime_ideal}
    Let $V\subseteq \Bbbk^n$ be an affine variety. $V$ is irreducible if and only if $\up{\tbf{I}}(V)$ is a prime ideal.
\end{proposition}

The Zariski Closure of generic subsets satisfy: 
\begin{lemma}
    If $W$ is a generic subset of $\Bbbk^k$, $\overline{W}=\Bbbk^k$.
    \label{lem:Zariski_Closure_of_Generic_Set}
\end{lemma}
\begin{proof}
    % We start by showing that $\Bbbk^k$ is irreducible.
    % $\Bbbk \in\{\R,\C\}$ is an infinite field. 
    Clearly $\bI(\Bbbk^k) = \{0\}$; this can be trivially shown to be a prime ideal.
    Then, by Proposition \ref{prop:irreducible_prime_ideal}, $\Bbbk^k$ is irreducible.
    %$\forall fg \in \bI(\Bbbk^k)$, $fg = 0$, so either $f = 0$ (i.e. $f\in\bI(\Bbbk^k)$) or $g = 0$ (i.e. $g\in\bI(\Bbbk^n)$), which shows $\bI(\Bbbk^k)$ is a prime ideal and thus $\Bbbk^k$ is irreducible.\\

    Since $W$ is a generic subset, we can write 
    $\Bbbk^k\backslash\bV(I) \subseteq W$ for some ideal nonzero ideal $I$.
    
    % \begin{equation}
    %     \begin{aligned}
    %         W = \Bbbk^k\backslash\bV(I)\subseteq \overline{\Bbbk^k\backslash\bV(I)} 
    %     \end{aligned}
    % \end{equation}
    % We have

    By definition of the Zariski Closure, we have   
    \begin{equation}
        (\Bbbk^k\backslash\bV(I))\cup\bV(I) \subseteq \overline{\Bbbk^k\backslash\bV(I)}\cup\bV(I)
    \end{equation}
    The left-hand side is equal to $\Bbbk^k$.
    For the right-hand side, since $\overline{\Bbbk^k\backslash\bV(I)} \subseteq \Bbbk^k$ and $\bV(I) \subseteq \Bbbk^k$, we  can conclude that the right-hand side is also equal to $\Bbbk^k$, i.e.,
    \begin{equation}
        \overline{\Bbbk^k\backslash\bV(I)}\cup \bV(I) = \Bbbk^k
    \end{equation}
    Since $I$ is nonzero, $\bV(I)\neq\Bbbk^k$. 
    By irreducibility of $\Bbbk^k$, $\overline{\Bbbk^k\backslash\bV(I)} = \Bbbk^k$.
    Then, by the properties of the Zariski Closure and definition of the generic subset, $\overline{\Bbbk^k\backslash\bV(I)} \subseteq \overline{W}\subseteq \Bbbk^k$, which gives $\overline{W} = \Bbbk^k$.
\end{proof}

We are now ready to prove Lemma~\ref{lem:Zariski_Closure_of_Generic_Subsets}.
\begin{proof}[Proof of Lemma~\ref{lem:Zariski_Closure_of_Generic_Subsets}]
    %When $I\neq\{0\}$, $\bV(I)\neq\Bbbk^n$. \\
    Proof by contradiction. Suppose that there exists $I\neq \{0\}$ such that $W\subseteq\bV(I)$. By Lemma~\ref{lem:Zariski_Closure_of_Generic_Set}, $\overline{W} = \Bbbk^k$. By the definition of Zariski Closure, $\overline{W}\subseteq \bV(I)$, which implies $\bV(I) = \Bbbk^k$. However, this can only be true if $I = \{0\}$, which contradicts the premise.
\end{proof}

% Lemma~\ref{lem:Zariski_Closure_of_Generic_Subsets} implies that a generic subset of an affine space is not contained in any hypersurface. We will be using this property in the later part of the section to determine the realization capacity.

\subsection{Controller parameterization with modal realization} \label{sec:controller_parameterization_a}

As stated in Section \ref{sec:equivalence}, to find a realization of a given strictly proper controller $\mbf{K}$ requires us to start with a state-space realization of the controller $\Sigma=(A,B,C,0)$ (see Lemma \ref{lem:realizability_and_T} for instance).
To facilitate analysis across different controllers, we first consider $\mbf{K}$ with $n$ distinct poles $p_i$, which can be expressed as $\mbf{K} = \sum_{i=1}^n\frac{1}{s-p_i}\Phi_i$ where $\Phi_i\in\C^{m\times q}$. 
For such controllers, we can always construct a modal realization:
\begin{equation}
\begin{aligned}
    A &= \begin{bmatrix}
        p_1I_m \\ & \ddots \\ & & p_nI_m
    \end{bmatrix},\quad  B = \begin{bmatrix}
        \Phi_1 \\ \vdots \\ \Phi_n
    \end{bmatrix}\\
    C &= \begin{bmatrix}
        I_m & \cdots & I_m
    \end{bmatrix}
\end{aligned}
\label{equ:MIMO_spectral_form}
\end{equation}

Collect parameters $p_i$ and $\Phi_i$ into vector $a\in\C^{n(mq+1)}$. Then, realization \eqref{equ:MIMO_spectral_form} can be re-written as:
\begin{equation}
\begin{aligned}
    A &= \begin{bmatrix}
        a_{nmq+1}I_m \\ & \ddots \\ & & a_{n(mq+1)}I_m
    \end{bmatrix}, \\
    B &= \begin{bmatrix}
        a_{1} & \cdots & a_{mn(q-1)+1}\\ \vdots & & \vdots \\ a_{nm} & \cdots & a_{nmq}
    \end{bmatrix}\\
    C &= \begin{bmatrix}
        I_m & \cdots & I_m
    \end{bmatrix}
\end{aligned}
\label{equ:MIMO_spectral_form_a}
\end{equation}

All strictly proper controllers $\mbf{K}$ of dimension $m\times p$ with $n$ distinct poles can be expressed in terms of $a\in\C^{n(mq+1)}$; we write $\mbf{K}(a)$ to make this explicit.
% We can then refer to parameterized modal realization \eqref{equ:MIMO_spectral_form_a} as:
% \begin{equation}
%     \Sigma_a = (A^a,B^a,C^a,0) \label{equ:parameterized_MIMO_realization_ta}
% \end{equation}
% and for a given controller $\mbf{K}(\ta)$, $a\in\C^{n(mq+1)}$, its modal realization is denoted by:
% \begin{equation}
%     \Sigma_{\ta} = (A^{\ta},B^{\ta},C^{\ta},0) \label{equ:short_MIMO_realization_ta}
% \end{equation}

\textbf{Remark:} The subsequent analysis assumes that we start with a modal realization \eqref{equ:MIMO_spectral_form_a}. 
These analyses trivially extend to any realization (e.g., canonical, Jordan); the only changes required are bookkeeping parameter collection into vector $a$.
Alternatively, controllers with repeated poles can be approximated by controllers with simple poles, which enables direct application of our analysis.

We also note that our assumption of $\mbf{K}$ being minimal is non-restrictive.
This is intuitive, and can in fact be stated rigorously in terms of generic subsets:

\begin{lemma}
    \label{lem:minimality_of_realization}
    Define the set of parameters
    \begin{equation}
        M_a := \{ a \in \C^{n(mq+1)} \mid \eqref{equ:MIMO_spectral_form_a} \up{ is a minimal realization}\}
    \end{equation}
    $M_a$ is a generic subset of $\C^{n(mq+1)}$.
\end{lemma}

\begin{proof}
Realization \eqref{equ:MIMO_spectral_form_a} is minimal if and only if the poles of the modal realization are distinct and each $\Phi_i$ in \eqref{equ:MIMO_spectral_form} are of full rank. The poles are distinct when:
    \begin{equation}
        \prod_{i\neq j}(p_i-p_j) \neq 0
    \end{equation}
    In \eqref{equ:MIMO_spectral_form_a} with parameters collected into $a$, this becomes:
    \begin{equation} \label{eq:distinct_pole_a}
        \prod_{i\neq j}(a_{nmq+i}-a_{nmq+j})\neq 0
    \end{equation}
    For $\Phi_i$, $i\in\{1,\cdots,n\}$ to have full rank, we require: 
    \begin{equation}
    \begin{aligned}
        \det(\Phi_i^*\Phi_i)\neq 0,\text{ } m\geq q\\
        \det(\Phi_i\Phi_i^*)\neq 0,\text{ } m<q
    \end{aligned}
        \label{equ:Phi}
    \end{equation}
    %\begin{itemize}
    %    \item when $m\geq q$, we have:
    %    \begin{equation}
    %        \det(\Phi_i^*\Phi_i)\neq 0 \label{equ:Phi_tall}
    %    \end{equation}
    %    \item When $m<q$, we have:
    %    \begin{equation}
    %        \det(\Phi_i\Phi_i^*)\neq 0 \label{equ:PhIide}
    %    \end{equation}
    %\end{itemize}
    We combine this with \eqref{eq:distinct_pole_a} and define the following polynomials $g_1, g_2$ in $\C[a]$ (note that $\Phi_i$ are also collected into vector $a$):
    \begin{equation}
        \begin{aligned}
            g_1\coloneq \prod_{i\neq j}(a_{nmq+i}-a_{nmq+j})\prod_{i=1}^n\det(\Phi_i^*\Phi_i) \\
            g_2\coloneq \prod_{i\neq j}(a_{nmq+i}-a_{nmq+j})\prod_{i=1}^n\det(\Phi_i\Phi_i^*)
        \end{aligned}
    \end{equation}
    
    When $m\geq q$, $g_1\neq 0$, $M_a = \C^{n(mq+1)}\backslash\bV(g_1)$.
    
    When $m < q$, $g_2\neq 0$, $M_a = \C^{n(mq+1)}\backslash\bV(g_2)$. 

    In both cases, $M_a$ satisfies the definition of a generic subset.
    
    %Therefore, for $\mbf{K}$ of arbitrary dimension, the set of $a$ which makes \eqref{equ:MIMO_spectral_form_a} a minimal realization is a generic subset of $\C^{n(mq+1)}$.
    %This means our assumption that $\mbf{K}$ is a minimal realization is not restrictive.
\end{proof}

\subsection{Generically realizing supports}

We are now ready to analyze Problem~\ref{pb:generically_realizing}, which is the problem of characterizing all controllers realizable on a given support. 
We first make this mathematically precise:
%With realization \eqref{equ:MIMO_spectral_form_a}, we can denote the set of controllers $\tbK=\mbf{K}(\ta)$ with simple poles that are realizable on a support $\Scal$, which is solution to Problem~\ref{pb:generically_realizing}, as:
\begin{definition}
    \label{def:realizable_set}
    The \textbf{realizable set} of a support $\Scal$ (as defined in Definition \ref{def:support}) is defined as:
    \begin{equation}
        \Rcal(\Scal)\coloneq\{\ta\in\C^{n(mq+1)}\mid \mbf{K}(\ta)\up{ is realizable on }\Scal\}
    \end{equation}
\end{definition}
% \begin{equation}
% \begin{aligned}
%     \tA &= \begin{bmatrix}
%         \ta_{nmq+1}I_m \\ & \ddots \\ & & \ta_{n(mq+1)}I_m
%     \end{bmatrix}, \\
%     \tB &= \begin{bmatrix}
%         \ta_{1} & \cdots & \ta_{mn(q-1)+1}\\ \vdots & & \vdots \\ \ta_{nm} & \cdots & \ta_{nmq}
%     \end{bmatrix}\\
%     \tC &= \begin{bmatrix}
%         I_m & \cdots & I_m
%     \end{bmatrix}
% \end{aligned}
% \label{equ:MIMO_spectral_form_ta}
% \end{equation}
In this section, we will analyze this realizable set.
Recall from Theorem~\ref{thm:problem_to_poly_eqns} that a controller $\mbf{K}$ with minimal realization $\tSigma=(\tA,\tB,\tC,0)$ is realizable on $\Scal$ if and only if we can find common zeros for all polynomials in $H_{\Scal,\tSigma, T_p}$. 
In this section, we introduce a change of notation, writing $H_{\Scal,\ta, T_p}$ instead of $H_{\Scal,\tSigma, T_p}$.
They represent the same polynomials, but the latter makes the dependence on controller parameters $\ta$ more explicit.
This lets us write the following definition:
\begin{definition}
    The \textbf{realizable set} of support $\Scal$ \textbf{under permutation matrix} $T_p\in\Pcal_p$ is defined as:
    \begin{equation}
        \Rcal(\Scal(T_p)) = \{\ta\in\C^{n(mq+1)}\mid H_{\Scal,\ta,T_p}\up{ has a common zero}\}
    \end{equation}
\end{definition}

These sets are related by the following:
\begin{lemma}
    \label{lem:decomposing_RS}
    The realizable set of support $\Scal$ is the union of realizable sets of $\Scal$ under all permutation matrices. i.e.,
    \begin{equation}
        \Rcal(\Scal) = \bigcup_{T_p\in\Pcal_p}\Rcal(\Scal(T_p))
        \label{equ:RS_decomposition}
    \end{equation}
\end{lemma}

We also introduce $H_{\Scal,a,T_p}$, which are symbolic polynomials whose coefficients depend on symbolic parameters $a$ (as opposed to fixed numerical values $\ta$ in $H_{\Scal,\ta, T_p}$).
We first make this technically precise:
% \begin{enumerate}
%     \item $H_{\Scal,a,T_p}$ is the set of polynomials corresponding to support $\Scal$, permutation matrix $T_p$ whose coefficients depend on the parameter $a$. 
%     \item For all $\ta\in\C^{n(mq+1)}$, $H_{\Scal,\ta,T_p}=\left.H_{\Scal, T_p ,a}\right|_{a = \ta}$. i.e. $H_{\Scal,\ta,T_p}$ is the polynomials corresponding to support $\Scal$, permutation matrix $T_p$ and controller $\mbf{K}(\ta)$.
% \end{enumerate}

\begin{definition}
    \label{equ:polyring_with_symbolic_coeff}
    The set of rational functions in $a$ is:
    \begin{equation}
        \begin{aligned}
            Q &\coloneq \{p = \frac{r(a)}{s(a)}\mid r(a),s(a)\in \C[a]\}
        \end{aligned}
    \end{equation}
    The set of polynomials with coefficients which are rational functions of $a$ is:
    \begin{equation}
        Q[x] \coloneq \{f(x) = \sum_\gamma p_\gamma x^\gamma,\quad p\in Q\}
    \end{equation}
\end{definition}

With this definition, we can write
\begin{equation}
\begin{aligned}
    \forall T_p\in\Pcal, \quad H_{\Scal,a,T_p}\subseteq Q[x] \\
    \forall T_p\in\Pcal, \quad \ta\in\C^{n(mq+1)}, \quad H_{\Scal,\ta,T_p}\subseteq \C[x]
\end{aligned}
\end{equation}

Different supports have different realizable sets. 
We now provide a small example to demonstrate this.
We consider $m=q=1$, $n=2$. Possible permutations are: 
\begin{equation}
    \Pcal_p = \{I_2,\begin{bmatrix}
        0 & 1 \\ 1 & 0
    \end{bmatrix}\}
\end{equation}

We first consider support $\Scal_1 = (\Acal_1,\Bcal_1,\Ccal_1,0)$, with
\begin{equation}
    \label{equ:S1}
    \Acal_1 = \begin{bmatrix}
        1 & 1 \\ 1 & 1
    \end{bmatrix}, 
    \Bcal_1 = \begin{bmatrix}
        1 \\ 0
    \end{bmatrix},
    \Ccal_1 = \begin{bmatrix}
        1 & 0
    \end{bmatrix}
\end{equation}

When $T_p = I_2$, we have
% $H_{\Scal_1,a,T_p} = \{-a_3x_1+a_4, x_1x_2+x_2+1\}$ is:
\begin{equation}
\begin{aligned}
H_{\Scal_1,a,T_p} = \{-a_3x_1+a_4, \quad x_1x_2+x_2+1\}
\end{aligned}
\label{equ:HTa1}
\end{equation}

A common zero exists for these polynomials when:
\begin{enumerate}
    \item $a_3 = 0$ and $a_4\neq 0$ do not hold simultaneously.
    \item $a_3+a_4\neq 0$.
\end{enumerate}
This implies that
\begin{equation}
    \C^{4}\backslash\bV(a_3(a_3+a_4))\subseteq \Rcal(\Scal_1(I_2))
\end{equation}

Since $\Rcal(\Scal_1(I_2))\subseteq \Rcal(\Scal_1)$, $\Rcal(\Scal_1)\supseteq \C^{4}\backslash\bV(a_3(a_3+a_4))$ is a generic subset of $\C^{4}$. 
In other words, the realizable set of $\Scal_1$ is a generic subset and includes nearly all possible parameters $a$; that is, \tbf{nearly all controllers can be realized on $\Scal_1$}.

We next consider support $\Scal_2 = (\Acal_2,\Bcal_2,\Ccal_2,0)$ where
\begin{equation}
    \Acal_2 = \begin{bmatrix}
        1 & 1 \\ 0 & 1
    \end{bmatrix}, 
    \Bcal_2 = \begin{bmatrix}
        0 \\ 1
    \end{bmatrix},
    \Ccal_2 = \begin{bmatrix}
        1 & 0
    \end{bmatrix}
    \label{equ:S2}
\end{equation}

When $T_p = I_2$, we have
\begin{equation}
    \begin{aligned}
    H_{\Scal_2,a,T_p} = \{(a_2-a_1)x_1,\quad x_1x_2+x_2+1, \\a_3x_1x_2-a_4x_2+a_3\}
\end{aligned}
\end{equation}

No common zeros exist for these polynomials unless $a_3 +a_4 = 0$, i.e., $\Rcal(\Scal_2(T_p))=\bV(a_3+a_4)$.

With $T_p = \begin{bmatrix}
    0 & 1 \\ 1 & 0
\end{bmatrix}$, we have
\begin{equation}
    \begin{aligned}
        H_{\Scal_2,a,T_p} = \{(a_1-a_2)x_1,\quad 
        x_1x_2+x_2+1,\\
        a_4x_1x_2-a_3x_2+a_4\}
    \end{aligned}
\end{equation}

Which also yields $\Rcal(\Scal_2(T_p))=\bV(a_3+a_4)$. Thus, based on Lemma \ref{lem:decomposing_RS}, $\Rcal(\Scal_2) = \bV(a_3+a_4)$.
This realizable set excludes most parameters $a$; that is,\textbf{ very few controllers can be realized on} $\Scal_2$.
% By Lemma~\ref{lem:Zariski_Closure_of_Generic_Set}, $\Rcal(\Scal_2)$ is not a generic subset of $\C^{4}$. 
In this example, one support can realize nearly all controllers.
This motivates the following definition:
%The different nature of $\Rcal(\Scal_1)$ and $\Rcal(\Scal_2)$ motivates the following definition:

\begin{definition}
    A support $\Scal$ is \tbf{generically realizing} if $\Rcal(\Scal)$ is a generic subset of $\C^{n(mq+1)}$.
\end{definition}

In our examples, $\Scal_1$ is generically realizing whereas $\Scal_2$ is not. 
$\Scal_2$ can realize very few controllers; in fact all supports that are not generically realizing also share this property.
This is shown later in Corollary \ref{cor:not_gen_realizing}.

Some generically realizing supports $\Scal$ satisfy $\Rcal(\Scal) = \C^{n(mq+1)}$. Those can realize \textit{all} controllers under parameterization \eqref{equ:MIMO_spectral_form_a}. 
One example of such a support is the support of the modal realization.
We provide another example below.

\begin{theorem}
    For single-input-single-output systems ($m=q=1$) with $n$ states, a generically realizing support $\Scal = (\Acal,\Bcal,\Ccal,0)$ with $\Rcal(\Scal) = \C^{2n}$ is:
    \begin{equation}
    \begin{aligned}
        \Acal_{ij} &= 1 \text{ if $i=1$, $j=1$ or $i=j$,} & \text{$0$ otherwise} \\
        \Bcal_{ij} &= 1 \text{ if $i=1$ or $i=n$,} &\text{$0$ otherwise} \\
        \Ccal_{ij} &= 1 \text{ if $j=1$ or $j=n$,} &\text{$0$ otherwise} \\
    \end{aligned}
    \end{equation}
\end{theorem}    
\begin{proof}
    We show that for any controller, we can construct an appropriate similarity transform to convert its modal realization to a realization satisfying $\Scal$ (akin to Corollary \ref{thm:problems_equal_finding_T}).
    First, note that compatible controllers can be written as
    \begin{equation}
        \mbf{K}(\ta) = \sum_{i=1}^n\frac{\ta_i}{s-\ta_{n+i}}
    \end{equation}
    Choose indices so that $\ta_{n}\neq 0$. $\mbf{K}(\ta)$ has modal realization:
    \begin{equation}
    \begin{aligned}
        \tA_m &= \mathrm{diag}(\ta_{n+1}, \dots,\ta_{2n}), \text{ }
        & \tB_m &= \begin{bmatrix}
            \ta_1 & \cdots & \ta_n
        \end{bmatrix}^\top,\\
        \tC_m &= \begin{bmatrix}
            1 & \cdots & 1
        \end{bmatrix}
    \end{aligned}
    \end{equation}
    Define $T_0 \coloneq\prod_{i=1}^{2(n-1)}E_i$ with 
    \begin{equation}
    \begin{aligned}
        E_i &= \begin{cases}
            E_{1,i+1}^{a}(\tx_i) = I_n+\tx_i\delta_{n,i+1}, &i \leq n-2\\
            E_{i+2-n,n}^{a}(\tx_i) = I_n+\tx_i\delta_{i+1,1}, &i \geq n-1
        \end{cases} \\
        \tx_i &= \begin{cases}
            -1,&\quad i = 1,\cdots,n-2 \\
            \frac{\ta_{i+2-n}}{\ta_n},&\quad i = n-1,\cdots,2(n-2)
        \end{cases}
    \end{aligned}
    \end{equation}
    Then, realization $(T_0\tA_mT_0^{-1},T_0\tB_m,\tC T_0^{-1},0)$ satisfies $\Scal$.
\end{proof}

We now introduce one of our main results: a theorem to determine whether a support is generically realizing.

\begin{theorem}
\label{thm:properties_of_generically_realizing}
    A support $\Scal$ is generically realizing if and only if there exists a permutation matrix $T_p\in \Pcal_p$ such that the reduced Gr\"obner basis of the ideal generated by $H_{\Scal,a,T_p}$, denoted by $G_{\Scal,a,T_p}$, satisfies:
    \begin{equation}
        G_{\Scal,a,T_p}\neq \{1\}
    \end{equation}
    Here, the reduced Gr\"obner basis of the ideal generated by $H\subseteq Q[x]$ is computed with the same algorithm used for polynomials in $\C[x]$, with every nonzero function of $a$ treated as a nonzero coefficient.
\end{theorem}

To prove Theorem~\ref{thm:properties_of_generically_realizing} requires analysis of the computations involved in obtaining the symbolic reduced Gr\"obner basis.
This requires the following definition and Lemma.

\begin{definition}
    Let $f,g\in \C[x]$ be nonzero polynomials, with $\alpha = \up{multidegree}(f)$ and $\beta=\up{multidegree}(g)$. Let $\gamma_i \coloneq \up{max}(\alpha_i,\beta_i)$, and $\gamma\coloneq(\gamma_1,\cdots,\gamma_n)$. The \tbf{S-polynomial} of $f$ and $g$ is defined as:
        \begin{equation}
            S(f,g) \coloneq \frac{x^\gamma}{\up{LT}(f)}\cdot f - \frac{x^\gamma}{\up{LT}(g)}\cdot g \label{equ:S_polynomial}
        \end{equation}
    where $x^\gamma$ and $\up{LT}()$ are as in Definitions \ref{def:degree} and \ref{def:multdeg_and_LT}.
\end{definition}

\begin{definition}
    Given $f \in Q[x]$ and $H = \{h_1,\cdots,h_l\}\subseteq Q[x]$ where $h_1\succ h_2\succ \cdots h_l$ with respect to monomial order $\succ$,  $\overline{f}^H$ is the \tbf{remainder} obtained by dividing $f$ by $h_1,h_2,\cdots,h_l$ sequentially.
\end{definition}

\begin{lemma}
    $\Rcal(\Scal(T_p))$ is a generic subset of $\C^{n(mq+1)}$ if and only if $G_{\Scal,a,T_p}\neq \{1\}$, where $G_{\Scal,a,T_p}$ is the reduced Gr\"obner basis of the ideal generated by $H_{\Scal,a,T_p}$. 
    % Moreover, when $G_{\Scal,a,T_p} = \{1\}$, there exists $r_{T_p}\in\C[a]$ such that $\Rcal(\Scal(T_p))\subseteq\bV(r_{T_p})$.
    \label{lem:generic_subset_equiv}
\end{lemma}

\begin{proof}
    Assume there exists a generic subset $W \subseteq \C^{n(mq+1)}$, where $\forall \ta \in W, G_{\Scal,\ta,T_p}$ is the reduced Gr\"obner basis of the ideal generated by $H_{\Scal,\ta,T_p}$. If such a $W$ exists, then:
    
    \textit{Sufficiency.} Since $G_{\Scal,a,T_p} \neq \{1\}$, $G_{\Scal,\ta,T_p} \neq 1, \forall \ta \in W$. Then, by Proposition~\ref{prop:weak_nullstellensatz}, solutions to $H_{\Scal,\ta,T_p}$ exist, and by Theorem~\ref{thm:problem_to_poly_eqns} and by definition of the realizable set, $\ta \in \Rcal(\Scal(T_p))$. This implies that $W \subseteq \Rcal(\Scal(T_p))$, which implies that $\Rcal(\Scal(T_p))$ is a generic subset.    

    \textit{Necessity.} We prove the contrapositive statement. When $G_{\Scal,a,T_p} = \{1\}$, $G_{\Scal,\ta,T_p} = \{1\}$ $\forall \ta \in W$.
    % where $\C^{n(mp+1)}\backslash\bV(r_{T_p})\subseteq W$. 
    Then, by Proposition~\ref{prop:weak_nullstellensatz}, solutions to $H_{\Scal,\ta,T_p}$ do not exist. By definition of the realizable set, $\forall \ta \in W$, $\ta \notin \Rcal(\Scal(T_p))$. This implies that $\Rcal(\Scal(T_p))\subseteq \C^{n(mq+1)}\backslash W$. Since $W$ is a generic subset, this implies that $\Rcal(\Scal(T_p))$ is not a generic subset. 

    The remainder of the proof shows that such a generic subset $W \subseteq \C^{n(mq+1)}$ exists by construction.
    To do so, notice that $G_{\Scal, a, T_p}$ can be obtained from $H_{\Scal, a, T_p}$ by applying Buchberger's Algorithm and subsequent Gr\"obner basis reduction \cite{cox1997ideals}, which consist of applying limited types of computations.
    Let $F$ denote the set of polynomials obtained at any intermediate step in the algorithms, and let $f_1, f_2 \in F$.
    Then, only the following manipulations are done to $F$:
    \begin{enumerate}
        \item Appending $\overline{S(f_1, f_2)}^F$ to $F$.
        \item Appending $\overline{f_1}^{F\backslash \{f_1\}}$ to $F$.
        \item Replacing $f_1$ in $F$ with $\frac{f_1}{\LC(f_1)}$.
        \item Removing polynomial $f_1$ from $F$.
    \end{enumerate}
    %In Buchberger's Algorithm and the subsequent Gr\"obner basis reduction \cite{cox1997ideals}, we use $F$ to denote the set of polynomials obtained at some intermediate step in the algorithm. With selected monomial order being $\succ$, computations involved fall into:
    %\begin{enumerate}
    %    \item Given $g_1,g_2\in F$, compute their $S-$polynomial:
    %    \begin{equation}
    %        g_{12} = S(g_1,g_2).
    %    \end{equation}
    %    \item Polynomial division and appending the remainder. Given $g = S(g_i,g_j)$, where $g_i,g_j\in F$, compute $\overline{f}^{F\backslash\{g\}}$. Then, if $\overline{f}^{F\backslash\{g\}}\neq 0$, update $F$ to be $F\cup \{\overline{f}^{F\backslash\{g\}}\}$.
    %    \item Dividing a polynomial by its leading coefficient, the coefficient of the polynomial's leading term. Given $g\in F$, update $F$ to be $(F\backslash\{g\})\cup \frac{g}{\LC(g)}$
    %    \item Removing a polynomial. Given $g\in F$, update $F$ to be $F\backslash \{g\}$.
    %\end{enumerate}
    Initially, $F = H_{\Scal,a,T_p}\subseteq Q[x]$.
    All four types of manipulations preserve this property; therefore, ultimately, $F \subseteq Q[x]$ at any step in the algorithm, and at termination, $F = G_{\Scal,a,T_p}\subseteq Q[x]$.   
    %\begin{itemize}
    %    \item For 2), if $g_i$, $g_j\in Q[x]$, $g\in Q[x]$. When $g\in Q[x]$, $F\backslash\{g\}\subseteq Q[x]$, $g^{H\backslash\{g\}}\in Q[x]$. 
    %    \item For 3), if $g\in Q[x]$, $\LC(g)\in S$, then $\frac{g}{\LC(g)}\in Q[x]$.
    %\end{itemize}
    Recall that we are trying to construct $W \subseteq \C^{n(mq+1)}$ such that $\forall \ta \in W, G_{\Scal,\ta,T_p}$ is the reduced Gr\"obner basis of the ideal generated by $H_{\Scal,\ta,T_p}$.
    The only case when $G_{\Scal,\ta,T_p}$ is \textit{not} the reduced Gr\"obner basis of the ideal generated by $H_{\Scal,\ta,T_p}$ is when division by zero (due to the value of $\ta$) occurs in one of the four types of manipulations.
    Thus, we now identify such cases of $\ta$ and exclude them from $W$ by construction.
    Division by zero will never occur for 4). We analyze the others.
    First, for all polynomials $f_i$ in $F$, $i= 1,\ldots,\zeta$, we write the leading coefficients in terms of numerators and denominators in $\mathbb{C}[a]$: $\LC(f_i)\coloneq \frac{n_i(a)}{d_i(a)}$. Then,
    \begin{enumerate}
        \item Division by zero does not occur for an S-polynomial computation 
        %as long as:
        %\begin{equation}
        %    \ta\in\C^{n(mq+1)}\backslash \bV(n_1(a)n_2(a)d_1(a)d_2(a))
        %\end{equation}
        or subsequent polynomial division when 
        \begin{equation}
            \ta\in\C^{n(mq+1)}\backslash\bV(\prod_{i=1}^\zeta n_i(a)d_i(a))
            \label{equ:manipulation1}
        \end{equation}
        
        \item Division by zero does not occur for this manipulation as long as:
        \begin{equation}
            \ta\in \C^{n(mq+1)}\backslash \bV(\prod_{i=2}^{\zeta} n_i(a)d_i(a))
            \label{equ:manipulation2}
        \end{equation}
        \item Division by zero does not occur for this manipulation as long as:
        \begin{equation}
            \ta\in\C^{n(mq+1)}\backslash\bV(n_1(a)d_1(a))
            \label{equ:manipulation3}
        \end{equation}
    \end{enumerate}
    %Note that for arbitrary polynomials $h_1,h_2\in\C[a]$:
    %\begin{equation}
    %\begin{aligned}
        %(\C^{n(mq+1)}\backslash\bV(h_1))\cap(\C^{n(mq+1)}\backslash\bV(h_2)) \\= \C^{n(mq+1)}\backslash\bV(h_1h_2)
    %\end{aligned}
    %\end{equation}
    Overall, $G_{\Scal, a, T_p}$ is obtained from $H_{\Scal, a, T_p}$ through a finite number of manipulations 1)-4). Therefore, we can collect their associated conditions into a single condition:
    \begin{equation}
        \ta \in \C^{n(mq+1)}\backslash \bV(w_{\Scal,T_p}(a))
    \end{equation}
    where $w_{\Scal,T_p}(a)\in\C[a]$ is constructed by collecting relevant terms from \eqref{equ:manipulation1}-\eqref{equ:manipulation3}.
    Then, we can select $W = \C^{n(mq+1)}\backslash \bV(w_{\Scal,T_p}(a))$, which is a generic subset with the desired properties.    
    % Then, for all $\ta\in \C^{n(mq+1)}\backslash \bV(u_{\Scal,T_p})$, when $G_{\Scal,a,T_p} = \{1\}$, $G_{\Scal,\ta,T_p} = \{1\}$, implying a solution does not exist for $H_{\Scal,\ta,T_p}$ and $\ta\notin\Rcal(\Scal(T_p))$. Therefore:
    % \begin{equation}
    %     \Rcal(\Scal(T_p))\subseteq \bV(u_{\Scal,T_p})
    % \end{equation}
    % By Lemma~\ref{lem:Zariski_Closure_of_Generic_Subsets}, $\Rcal(\Scal(T_p))$ is not a generic subset of $\C^{n(mq+1)}$.

    % For all $\ta\in \C^{n(mq+1)}\backslash \bV(u_{\Scal,T_p})$, when $G_{\Scal,a,T_p} \neq \{1\}$, $G_{\Scal,\ta,T_p} \neq \{1\}$, implying a solution exists for $H_{\Scal,\ta,T_p}$ and $\ta\in\Rcal(\Scal(T_p))$. Therefore:
    % \begin{equation}
    %     \Rcal(\Scal(T_p))\supseteq \C^{n(mq+1)}\backslash\bV(u_{\Scal,T_p})
    % \end{equation}
    % By Lemma~\ref{lem:Zariski_Closure_of_Generic_Subsets}, $\Rcal(\Scal(T_p))$ is a generic subset of $\C^{n(mq+1)}$.
\end{proof}

With this, we are now ready to prove Theorem~\ref{thm:properties_of_generically_realizing}.

\begin{proof}[Proof of Theorem~\ref{thm:properties_of_generically_realizing}]
Recall from Lemma~\ref{lem:decomposing_RS}:
\begin{equation}
    \Rcal(\Scal) = \bigcup_{T_p\in\Pcal_p}\Rcal(\Scal(T_p))
    \label{equ:Rcal_Scal}
\end{equation}

\textit{Sufficiency}: When there exists $T_p\in\Pcal_p$ such that $G_{\Scal,a,T_p}\neq \{1\}$, by Lemma \ref{lem:generic_subset_equiv}, $\Rcal(\Scal(T_p))$ is a generic subset. Then, there exists $w_{\Scal,T_p}(a)\in\C[a]$ such that $\C^{n(mq+1)}\backslash\bV(w_{\Scal,T_p}(a)) \subseteq \Rcal(\Scal(T_p)) \subseteq \Rcal(\Scal)$, so $\Rcal(\Scal)$ is a generic subset of $\C^{n(mq+1)}$.

\textit{Necessity}: We will prove the contrapositive. Assume $G_{\Scal,a,T_p} = \{1\}$ for all $T_p\in\Pcal_p$. Then, by Lemma \ref{lem:generic_subset_equiv} and results from its proof, for all $T_p$ there exists a generic subset $W = \C^{n(mq+1)}\backslash\bV(w_{\Scal,T_p}(a))$ such that $G_{\Scal,\ta,T_p} = \{1\}$ for all $\ta\in W$. 
This means $\Rcal(\Scal(T_p))\subseteq\bV(w_{\Scal,T_p}(a))$. Then, $\Rcal(\Scal) \subseteq \bigcup_{T_p\in\Pcal_p}\bV(w_{\Scal,T_p}(a)) = \bV(\prod_{T_p\in\Pcal_p}w_{\Scal,T_p}(a))$. So $\Rcal(\Scal)$ is not a generic subset of $\C^{n(mq+1)}$.%, which proves the theorem.

% $\Scal$ is generically realizing if and only if
% \begin{equation}
%     \overline{\Rcal(\Scal)} = \bigcup_{T_p\in\Pcal_p}\overline{\Rcal(\Scal(T_p))} = \C^{n(mq+1)}
%     \label{equ:decomposing_bRS}
% \end{equation}
% Proof of Lemma~\ref{lem:Zariski_Closure_of_Generic_Set} shows that $\C^{n(mq+1)}$ is irreducible. \eqref{equ:decomposing_bRS} contains a finite union. Thus, $\Scal$ is generically realizing if and only if there exists $T_p\in\Pcal_p$ such that:
% \begin{equation}
%     \overline{\Rcal(\Scal(T_p))} = \C^{n(mq+1)}
% \end{equation}

% Then, by Lemma~\ref{lem:generic_subset_equiv}, $\overline{\Rcal(\Scal(T_p))} = \C^{n(mq+1)}$ if and only if $G_{\Scal,a,T_p}\neq\{1\}$, which proves the theorem.
\end{proof}    

% For support $\Scal_2$, which is not generically realizing, $\C^{n(mq+1)}\backslash \Rcal(\Scal_2)$ is a generic subset of $\C^{n(mq+1)}$. This holds for all supports that are not generically realizing; only a small set of controllers are realizable on such supports. That is, 

The proof of this theorem also gives the following result:

\begin{corollary}
    If a support $\Scal$ is not generically realizing, then there exists some polynomial $w(a)\in\C[a]$ such that:
    \begin{equation}
        \Rcal(\Scal)\subseteq \bV(w(a))
    \end{equation}
    \label{cor:not_gen_realizing}
\end{corollary}
This completes our main result in this section: namely, any given support is either generically realizing (can realize most controllers) or can realize few controllers (only those whose parameters belong to a variety).
If we consider a controller $\mbf{K}(\ta)$ with randomly generated parameters $\ta$, the probability of $\ta$ belonging to a given variety is nearly zero.
This means that if $\Scal$ is generically realizing, it can almost certainly realize $\mbf{K}(\ta)$; if it is not, then it can almost certainly not realize $\mbf{K}(\ta)$.

The following corollary shows that for single-input single-output (SISO) systems, we only need to consider $T_p = I$.

\begin{corollary}
    A support $\Scal$ of a SISO system is generically realizing if and only if $G_{\Scal,a,I}\neq \{1\}$. 
\end{corollary}

\begin{proof}
    For SISO systems, applying permutation $T_p$ to modal realization \eqref{equ:MIMO_spectral_form_a} permutes rows in $B$ and diagonal terms in $A$. Then, for all $T_p\in\Pcal_p$, there exists some permutation of variable $a$, denoted by ${\hat{a}}$ such that $H_{\Scal,a,T_p} = H_{\Scal,\hat{a},I}$.
    % By Theorem~\ref{thm:properties_of_generically_realizing}, a support is generically realizing if and only if there exists $T_p\in\Pcal_p$ such that the reduced Gr\"obner basis of $H_{\Scal,a,T_p}$, denoted by $G_{\Scal,a,T_p}$ does not equal $\{1\}$. 
    % Define $\hat{a}$ as a permutation of the variables in $a$. For instance, if $a = [a_1,a_2]^\top$, a valid $\hat{a}$ is $\hat{a} = [a_2,a_1]^\top$. For SISO systems, for all $T_p\in\Pcal_p$, there exists $\hat{a}$, a permutation of $a$ such that $H_{\Scal,T_a,a}\subseteq Q[x]$ satisfies:
    % \begin{equation}
    %     H_{\Scal,T_a,a} = H_{\Scal,I_n,\hat{a}}
    % \end{equation}
    This implies that $G_{\Scal,a,T_p} = G_{\Scal,\hat{a},I}$.
    Since $\hat{a}$ is a permutation of $a$, $G_{\Scal,\hat{a},I}\neq\{1\}$ if and only if $G_{\Scal,a,I}\neq 1$. 
    Applying Theorem~\ref{thm:properties_of_generically_realizing} proves the claim.
    % For SISO systems, with a symbolic modal realization $(A,B,C,0)$ as shown in \eqref{equ:MIMO_spectral_form_a} with parameters $a$, for any $T_p\in\Pcal_p$, $(T_p^{-1}AT_p,T_p^{-1}B,CT_p,0)$ becomes 
    % Then, for any other $T_p$, since the system is SISO, we now obtain $\tilde{F} = \left.F_0\right|_{a = \tilde{a}}$ and $\tilde{G} = \left.G_0\right|_{a = \tilde{a}}$, where $\tilde{a}$ is permutation of $a$. It then follows that $\tilde{G}\neq\{1\}$, which proves the claim.
\end{proof}

% The proof of Theorem~\ref{thm:properties_of_generically_realizing} reveals a property of non generically realizing supports: a support can realize either nearly all controllers or only a very limited set of controllers. If a support $\Scal$ is generically realizing, when we apply a sufficient sampling of controller parameters. The vast majority, if not all of the parameters lie in $\Rcal(\Scal)$. If a support is not generically realizing, when we apply a sufficient sampling of different controller parameters. The vast majority, if not all the parameters does not lie in $\Rcal(\Scal)$. As computing Gr\"obner basis with numbers is less time consuming than computing Gr\"obner basis with parameter $a$, in the next subsection, we leverage this property, along side other criteria to determine if a support $\Scal$ is generically realizing.

% Thus, if a controller with randomly generated parameters can be realized on a support $\Scal$, it is highly likely that $\Scal$ is generically realizing. This can be desirable as computing Gr\"obner basis with numbers is faster than computing Gr\"obner basis with symbolic variables. Algorithm~\ref{alg:near_gen_realizing} proposed at the end of the section leverages these properties. 

\subsection{Necessary conditions for generically realizing}
\label{sec:necessary_criteria}

Using Theorem~\ref{thm:properties_of_generically_realizing}, we can determine whether a support is generically realizing by computing reduced Gr\"obner bases.
This is computationally involved.
In this subsection, we introduce some necessary conditions for a support to be generically realizing; these can be used for efficient screening.

\begin{theorem}
    If $\Scal$ is generically realizing, $\exists i$ such that the $i$-th row of $\Bcal$ and the $i$-th column of $\Ccal$ are both nonzero.
    \label{thm:io_connection}
\end{theorem}
\begin{proof}
    We prove the contrapositive. Assume that $\forall i \in \{1, \ldots, n\}$, either the $i$-th row of $\Bcal$ or $i$-th column of $\Ccal$ is zero (or both). Then, for all realizations satisfying $\Scal$, we can rearrange the states to group nonzero inputs and outputs:
    \begin{equation}
        A = \begin{bmatrix}
            A_{11} & A_{12} \\ A_{21} & A_{22}
        \end{bmatrix},\quad
        B = \begin{bmatrix}
            0 \\ B_{21}
        \end{bmatrix},\quad
        C = \begin{bmatrix}
            C_{11} & 0
        \end{bmatrix}
    \end{equation}
    Write $(sI-A)^{-1}\coloneq \begin{bmatrix}
        \Psi_{11}(s) & \Psi_{12}(s) \\ \Psi_{21}(s) & \Psi_{22}(s)
    \end{bmatrix}$.
    The resulting controller transfer function is
        $\mbf{K} = C_{11}\Psi_{12}(s)B_{21}$.
    By properties of the matrix inverse, we have
    \begin{equation}
        s\Psi_{12}(s)-(A_{11}\Psi_{12}(s)+A_{12}\Psi_{22}(s)) = 0
    \end{equation}
    Since $A_{11}\Psi_{12}(s)+A_{12}\Psi_{22}(s)$ is zero or strictly proper, $s\Psi_{12}(s)$ is zero or strictly proper. This means that if $\Psi_{12}(s)\neq 0$, the relative degree of $\Psi_{12}(s)$ is at least 2, so we yield a transfer function matrix with relative degree at least $2$. The modal realization  \eqref{equ:MIMO_spectral_form} and associated partial fraction form implies that all systems with relative degree 2 or higher satisfy:
    \begin{equation}
       \sum_{i=1}^{n} \Phi_i = 0 \label{equ:sum_phi_i}
    \end{equation}
    Equation \eqref{equ:sum_phi_i} is an equality constraint on controller parameters $a$. 
    % We can come up with equations of a format similar to \eqref{equ:sum_phi_i} for all $T_p\in\Pcal_p$. 
    This means that there exists $w\in\C[a]$ such that $\Rcal(\Scal)\subseteq\bV(w)$, 
    so $\Scal$ is not generically realizing. 
\end{proof}

Structural controllability \cite{lin1974structural,shields2003structural} provides another necessary condition 
%Recall that a support is written as $\Scal=(\Acal,\Bcal,\Ccal,\Dcal)$.
\begin{definition}
    A support $\Scal$ is \tbf{structurally controllable} if there exists controllable $\tA, \tB$ satisfying this support. A support $\Scal$ is \textbf{structurally observable} if there exists observable $\tA, \tC$ satisfying this support.
\end{definition}
\begin{theorem}
    If $\Scal$ is generically realizing, then $\Scal$ is structurally controllable and structurally observable.
    \label{thm:gen_realizing_vs_struct_ctrb}
\end{theorem}
\begin{proof}
    We prove the contrapositive.
    Assume $\Scal$ is either not structurally controllable or not structurally observable. Then, any realization satisfying $\Scal$ cannot be minimal. 
    In other words, $\Rcal(\Scal) \subseteq N$ where 
    \begin{equation}
        N = \{\ta\mid \up{Realization \eqref{equ:MIMO_spectral_form_a} of $\mbf{K}(\ta)$ is not minimal}\}
    \end{equation}
    By Lemma~\ref{lem:minimality_of_realization}, $N$ is not a generic subset of $\C^{n(mq+1)}$, so $\Scal$ is not generically realizing.
\end{proof}

\subsection{Sufficient conditions for generically realizing}

In this subsection, we introduce sufficient conditions for a support to be generically realizing; this is more computationally efficient than computing reduced Gr\"obner bases as required by Theorem~\ref{thm:properties_of_generically_realizing}.

%To determine if $\Scal$ is generically realizing, we can check if $\Scal$ is accessible, which shows a support is generically realizing by decomposing invertible matrices in a different manner. Determining whether a support is accessible is less time consuming than computing Gr\"obner basis. We can also efficiently compute realizations of a given controller $\mbf{K}(\ta)$ that satisfies an accessible support $\Scal$.

First, a trivial definition and result.
\begin{definition}
    A support $\Scal' = (\Acal',\Bcal',\Ccal',\Dcal')$ is \tbf{denser} than support $\Scal = (\Acal,\Bcal,\Ccal,\Dcal)$ if:
    \begin{equation}
        \forall i,j,\up{ if $\mathcal{M}_{ij} = 1$, then $\mathcal{M}'_{ij} = 1$, }\mathcal{M}\in\{\Acal,\Bcal,\Ccal,\Dcal\}
    \end{equation}
    When $\Scal'\neq\Scal$, $\Scal'$ is \tbf{strictly denser} than $\Scal$. \textbf{Sparser} and \textbf{strictly sparser} are defined analogously.
\end{definition}

\begin{lemma}
    Let $\Scal$ be generically realizing. Then, any support $\Scal'$ that is denser than $\Scal$ is also generically realizing.
    \label{lem:denser_supps_are_gen_realizing}
\end{lemma}

Now, we begin analysis to provide a less trivial sufficient condition.
The idea is to find a sequence of elementary addition matrices to convert the modal realization to a realization satisfying the desired support.
Consider $\Sigma^*$, which is $\tSigma = (\tA,\tB,\tC,0)$ transformed by an elementary addition matrix $E_{ij}^a(x)$ \eqref{equ:elementary_addition}, i.e.,  
\begin{equation} \label{eq:sigma_star}
    \Sigma^* = ({E_{ij}^a(x)}^{-1}\tA E_{ij}^a(x), {E_{ij}^a(x)}^{-1}\tB, \tC E_{ij}^a(x),0)
\end{equation}
This realization contains symbols $x$; different numerical values of $x$ may lead $\Sigma^*$ to have different supports.
We define support ${\Scal^*} = ({\Acal^*},{\Bcal^*},{\Ccal^*},0)$ to be the densest possible support:
 \begin{equation} \label{eq:s_star}
        \begin{aligned}
            {\Acal^*}_{\mu\nu} &= \begin{cases}
                1 \quad &\up{$\Acal_{\mu\nu} = 1$} \\ 
                &\up{or $\nu=j$ and $\Acal_{\mu i}=1$}\\ 
                &\up{or $\mu=i$ and $\Acal_{\nu j} = 1$.}\\ 
                0 \quad &\up{otherwise}
            \end{cases} \\
            {\Bcal^*}_{\mu\nu} &= \begin{cases}
                1 \quad & \up{$\Bcal_{\mu\nu} = 0$}\\
                &\up{or $\mu=i$ and $\Bcal_{j\nu} = 1$} \\ 
                0 & \up{otherwise}
            \end{cases} \\
            {\Ccal^*}_{\mu\nu} &= \begin{cases}
               1 \quad & \up{$\Ccal_{\mu\nu} = 0$}\\
                &\up{or $\nu=j$ and $\Ccal_{\mu i} = 1$} \\ 
                0 & \up{otherwise}
            \end{cases}
        \end{aligned}
    \end{equation}
We now provide conditions relating $x$, $\Sigma^*$, and support $\Scal^*$.

\begin{lemma} \label{lem:densest_support_after_elementary_addition}
    Let $\Scal = (\Acal,\Bcal,\Ccal,0)$ denote the support of $\tSigma$, and let $\Sigma^*$ and $\Scal^*$ be as defined in \eqref{eq:sigma_star} and \eqref{eq:s_star}. 
    Then, there exists generic subset $U \subseteq\C$ such that for all $\tx\in U$, $\Scal^*$ is the support of $\tSigma^* = \Sigma^*\mid_{x = \tx}$. Furthermore, for all $\tx\in\C\backslash U$, the support of $\tSigma^*$ is strictly sparser than $\Scal^*$.
\end{lemma}

\begin{proof}    
    We can write $\Sigma^*=(A^*,B^*,C^*,0)$ as:
    \begin{equation}
    \begin{aligned}
        A^*_{\mu\nu} &= \begin{cases}
            \tA_{\mu j}+\tA_{\mu i}x,\quad& \mu\neq i, \nu = j \\
            \tA_{i\nu}-\tA_{j\nu}x,\quad& \mu = i,\nu\neq j \\
            \tA_{ij}+(\tA_{ii}-\tA_{jj})x-\tA_{ji}x^2 \\
            \quad& \mu = i, \nu = j\\
            \tA_{\mu\nu}, \quad&\up{otherwise}
        \end{cases} \\
        B^*_{\mu\nu} &= \begin{cases}
            \tB_{i\nu}-\tB_{j\nu}x,\quad & \mu = i\\
            \tB_{\mu\nu}, \quad & \up{otherwise}
        \end{cases} \\
        C^*_{\mu\nu} &= \begin{cases}
            \tC_{\mu j}+\tC_{\mu i}x,\quad & \nu = j\\
            \tC_{\mu\nu}, \quad &\up{otherwise}
        \end{cases}
    \end{aligned} 
    \label{equ:elementary_addition_1}
    \end{equation}
    Define nonzero product $u\in\C[x]$ as\footnote{    If $\tA_{ii}=0$, $\tA_{jj}=0$ and/or $\tA_{ji} = 0$, remove the term $\tA_{ij}+(\tA_{ii}-\tA_{jj})x-\tA_{ji}x^2$ from the product.}: 
    \begin{equation}
    \begin{aligned}
        u\coloneq &\prod_{\underset{\mu\neq i}{\tA_{\mu i}\neq 0}}(\tA_{\mu j}+\tA_{\mu i}x)\prod_{\underset{\nu\neq j}{\tA_{j\nu}\neq 0}}(\tA_{i\nu}-\tA_{j\nu}x)\\&(\tA_{ij}+(\tA_{ii}-\tA_{jj})x-\tA_{ji}x^2) \\
         & \prod_{\tB_{j\nu}\neq 0}(\tB_{i\nu}-\tB_{j\nu}x)\prod_{\tC_{\nu i}\neq 0}(\tC_{\nu j}+\tC_{\nu i}x)
    \end{aligned}
    \end{equation}
    Choose $U=\C\backslash\bV(u)$, which is a generic subset of $\C$. Then, for all $\tx\in U$, $\Scal^*$ is the support of $\tSigma^*$. When $\tx\in\C\backslash U$, one additional entry in one of $A^*$, $B^*$, $C^*$ is zero.
\end{proof}

The proof of Lemma~\ref{lem:densest_support_after_elementary_addition} provides a way for us to choose $\tx$ to set an entry in $A^*$, $B^*$ or $C^*$ to zero. To simplify analysis, we only consider the case where the entry is in $B^*$ or $C^*$.
Note that if we also consider $A^*$, the resulting analyses become much more involved, which would defeat the purpose of looking for a more efficient alternative to Theorem~\ref{thm:properties_of_generically_realizing}.

% We remove the edge in i with the edge in j
\begin{definition} \label{def:viable_conversion}
     $\Scal'=(\Acal',\Bcal',\Ccal',0)$ is a \tbf{viable conversion} of $\Scal$ under elementary addition matrix $E_{ij}^a(x)$ if 1) or 2) holds:
     \begin{enumerate}
    \item $\Acal'  = {\Acal^*}$, $\Ccal' = {\Ccal^*}$, and $\exists\tnu$ such that $\Bcal_{j\tnu} = {\Bcal^*}_{i\tnu}=1$ where
    \begin{equation}
    \begin{aligned}
        \Bcal'_{\mu\nu} = \begin{cases} 0 &\up{$\mu=i$, $\nu = \tnu$}
             \\ {\Bcal^*}_{\mu\nu} &\up{ otherwise}
        \end{cases}
    \end{aligned}
    \end{equation}
    \item $\Acal'  = {\Acal^*},\text{ } \Bcal' = {\Bcal^*}$, and $\exists\tmu$ such that $\Ccal_{\tmu i} = {\Ccal^*}_{\tmu j}=1$ where
    \begin{equation}
    \begin{aligned}
        \Ccal'_{\mu\nu} = \begin{cases} 0 &\up{$\mu=\tmu$, $\nu=j$}
             \\ {\Ccal^*}_{\mu\nu} &\up{otherwise}
        \end{cases}
    \end{aligned}
    \end{equation}
     \end{enumerate}
\end{definition}

$\Scal'$ is $\Scal^*$ with one entry in $\Bcal^*$ or $\Ccal^*$ set to zero. By \eqref{equ:elementary_addition_1}, in order to make $\Sigma^*$ satisfy $\Scal'$, depending on $\Bcal'$ and $\Ccal'$, one of the following two equalities must hold:
\begin{equation} \label{eq:viable_conversion}
    \begin{aligned}
        \tB_{i\nu} - \tB_{j\nu}x &= 0\\
        \tC_{\mu j} + \tC_{\mu i}x &= 0
    \end{aligned}
\end{equation}
% $\Scal^*$ is strictly denser than $\overline{\Scal}$ with an entry in $(E_{ij}^{-1}AE_{ij},E_{ij}^{-1}B,CE_{ij},D)$ that has value affine in $x$ set to zero. Practically speaking, $\overline{\Scal}$ is the support of the realization for the vast majority of $x$ and we can assign a value to $x$ to make the newer realization satisfy $\Scal^*$

We now introduce the main result of this subsection.
\begin{definition}
    \label{def:accessible}
    Let $\Scal^0$ be the support of any modal realization \eqref{equ:MIMO_spectral_form_a}, and let $\Scal = (\Acal,\Bcal,\Ccal,0)$ be a support with the same dimensions as $\Scal^0$; $\Scal$ is \tbf{accessible} if there exists a sequence $\{\Scal^l\}_{l=0}^k$ where $\Scal^l = (\Acal^l,\Bcal^l,\Ccal^l,0)$, such that:
    \begin{enumerate}
        \item For $l = 1,\ldots,k$, $\Scal^l$ is a viable conversion of $\Scal^{l-1}$.
        \item $\Scal^k$ is sparser than $\Scal$.
        \item If there exists $\mu,\nu$, $l$ such that $\Bcal_{\mu \nu}^l = 1$ and $\Bcal_{\mu \nu}^{l+1} = 0$, then $\Bcal_{\mu \nu}^{l+2} =\cdots = \Bcal_{\mu \nu}^k = 0$.
        \item If there exists $\mu$, $\nu$, $l$ such that $\Ccal_{\mu\nu}^l = 1$ and $\Ccal_{\mu\nu}^{l+1} = 0$, then $\Ccal_{\mu\nu}^{l+2} =\cdots = \Ccal_{\mu\nu}^k = 0$.
        \item For all $\mu$, $\nu$ such that $\Acal_{\mu\nu} = 0$, $\Acal_{\mu\nu}^0 = \Acal_{\mu\nu}^1 = \ldots = \Acal_{\mu\nu}^k = 0$ (i.e., $\Acal^l$ is sparser than $\Acal$).
    \end{enumerate}
\end{definition}

% Accessibility is a sufficient condition on generically realizing. 

\begin{theorem}
    If $\Scal$ is accessible, then $\Scal$ is generically realizing.
    \label{thm:accessible_supps_are_gen_realizing}
\end{theorem}
\begin{proof}
    By definition, we have a sequence of supports $\{\Scal^l\}_{l=0}^k$. 
    Any controller $\mbf{K}(a)$ has an associated realization $\Sigma^0=(A^0,B^0,C^0,0)$ which satisfies $\Scal^0$; this is simply the modal realization \eqref{equ:MIMO_spectral_form_a}.
    We now show that other realizations $\Sigma^l=(A^l,B^l,C^l,0)$ exist which satisfy supports $\Scal^l$ from the sequence; however, their existence requires additional polynomial equations to be satisfied.
    
    To construct $\Sigma^1$, apply elementary addition matrix $E_{i_1j_1}^a(x_1)$ to $\Sigma^0$ (similar to \eqref{eq:sigma_star}).
    For $\Sigma^1$ to satisfy $\Scal^1$, we require one of the following equations to hold, depending on the type of conversion that was applied to $\Scal^0$ to get $\Scal^1$:
    \begin{equation}
    \begin{aligned}
        B^0_{i_1\nu}-B^0_{j_1\nu} x_1&= 0\\
        C^0_{\mu j_1}+C^0_{\mu i_1} x_1 &= 0
    \end{aligned} 
    \label{equ:first_elementary_addition_constraints}
    \end{equation}
    Note the resemblance to \eqref{eq:viable_conversion}.
    Recall that $B^0$ and $C^0$ depend symbolically on controller parameters $a$. If there exists a solution to this equation, then $\Sigma^1$ is a realization of $\mbf{K}(a)$.
    Similar equations can be written (all resemble \eqref{eq:viable_conversion}) for $\Scal^2 ,\ldots,\Scal^k$. We collect these equations into:
    \begin{equation}
    \begin{aligned}
        f_{10}(a)+f_{11}(a)x_1 & = 0 \\
        f_{20}(a,x_1)+f_{21}(a,x_1)x_2 &= 0\\
        \vdots \\
        f_{k0}(a,x_1,\cdots,x_{k-1})+f_{k1}(a,x_1,\cdots,x_{k-1})x_k &= 0
    \end{aligned}       
    \label{equ:equation_generated_by_option_table}
    \end{equation}
    where $f_{l0},f_{l1}\in\C[a,x_1,\ldots,x_{l-1}] $ for $l=1,\ldots,k$.
    Here, $f_{l0}$ is nonzero due to 3) and 4) in Definition \ref{def:accessible} and $f_{l1}$ is nonzero due to Definition \ref{def:viable_conversion}.

    Now, assume there exists a generic subset $W \subseteq \C^{n(mq+1)}$, where $\forall \ta \in W$, solutions exist for \eqref{equ:equation_generated_by_option_table} with $a=\ta$. If such a $W$ exists, then for all $\ta \in W$, $\mbf{K}(\ta)$ is realizable on $\Scal^k$, so $W \subseteq \Rcal(\Scal^k)$; this means $\Scal^k$ is generically realizing. Then, by property 2) of Definition \ref{def:accessible} and Lemma~\ref{lem:denser_supps_are_gen_realizing}, $\Scal$ is generically realizing, which would conclude the proof.

    The remainder of the proof shows that such a $W$ exists by construction.
    %We claim that for equations of form \eqref{equ:equation_generated_by_option_table}, there exists a generic subset $W = \C^{n(mq+1)}\backslash\bV(u)$, $u\in\C[a]$ such that for all $\ta\in W$, a solution to \eqref{equ:equation_generated_by_option_table} has a solution of format $x_l = \frac{n_l(a)}{d_l(a)}$, where $n_l(a)$, $d_l(a)\in\C[a]$.
    First, consider $l=1$.
    Define $W_1 \coloneq \C^{n(mq+1)}\backslash\bV(f_{11}(a))$. Then, for all $\ta\in W_1$, i.e. $f_{11}(\ta)\neq 0$, $\tx_1 =-\frac{f_{10}(\ta)}{f_{11}(\ta)}$ solves the equation.
    Then, consider $l=2$. We plug in $x_1=\tx_1$ to obtain:
    \begin{equation} \label{eq:l2_tilde}
        f_{20}(a, \tx_1) + f_{21}(a, \tx_1)x_2 = 0
    \end{equation}
    Notice that now, $f_{20}, f_{21} \in Q$ (Definition \ref{equ:polyring_with_symbolic_coeff}).
    We rewrite the numerators and denominators explicitly:
    \begin{equation}
        \frac{n_{20}(a)}{d_{20}(a)}+\frac{n_{21}(a)}{d_{21}(a)}x_2 = 0
    \end{equation}
    where $n_{20}(a),n_{21}(a),d_{20}(a),d_{21}(a)\in\C[a]$. Define $W_2\coloneq\C^{n(mq+1)}\backslash\bV(d_{20}(a)n_{21}(a))$. Then, for all $\ta\in W_2$, $\tx_2 = -\frac{n_{20}(\ta)d_{21}(\ta)}{n_{21}(\ta)d_{20}(\ta)}$ solves the equation.
    For $l = 3,\ldots k$, apply the same procedure: plug in $\tx_1, \ldots \tx_{l-1}$ from previous equations, and define $W_l$ to avoid divisions by zero.
    Then, the set $W$ is the intersection of all $W_l$, i.e., $W \coloneq \bigcap_{l=1}^k W_l$.
    Notice that $W$ can be equivalently written as $\C^{n(mq+1)}\backslash\left(\bV(f_{11}(a) \cup \bV(d_{20}(a)n_{21}(a)) \cup \ldots \right)$ for a finite number of varieties, so it is a generic subset.
\end{proof}

\textbf{Remark:} If $\Scal$ is accessible, for $\ta \in W$, we can obtain a realization of $\mbf{K}(\ta)$ that satisfies $\Scal$ by solving for \eqref{equ:equation_generated_by_option_table}.

\subsection{Algorithm for generically realizing }

We now introduce Algorithm~\ref{alg:generically_realizing}, which determines whether $\Scal$ is generically realizing.
This algorithm first checks the two necessary conditions from subsection \ref{sec:necessary_criteria}, which are numerically cheap to compute.
Specifically, structural controllability and observability can be computed using a fixed-zero-root-finder \cite{shields2003structural}.
If these are satisfied, the algorithm checks whether $\Scal$ is accessible using a depth-first tree search; this is denoted as $\up{IsAccessible}(\Scal)$.
Here, each support is treated as a node on the tree; the root is $\Scal^0$.
For any support $\Scal^l$, the number of viable conversions is finite as there are only finite combinations of entries $i_l$ and $j_l$, and for a fixed $i_l$ and $j_l$, $\Scal^{l+1}$ can only take two distinct values; so the number of branches from any node is finite. 
Candidates for $\Scal^{l+1}$ can be found by generating all possible conversions of $\Scal^l$ satisfying properties 3)-5) in Definition~\ref{def:accessible}. 
Note that properties 3) and 4) in Definition~\ref{def:accessible} enforce that starting from $\Scal^l$, only a finite number of nonzero entries in $\Bcal^l$ and $\Ccal^l$ can be set to zero; so the depth of the tree is also finite. 
This search either yields the required sequence $\Scal^0,\ldots,\Scal^k$ or certifies inaccessibility.

If $\Scal$ is accessible, then it is generically realizing by 
Theorem~\ref{thm:accessible_supps_are_gen_realizing}. 
This is only a sufficient condition; if $\Scal$ is not accessible, we must apply Theorem~\ref{thm:properties_of_generically_realizing}, which requires computation of symbolic Gr\"obner bases.
Note that if speed is preferred over rigor, we could alternatively apply Corollary~\ref{cor:not_gen_realizing} and the remarks following it; that is, instead of computing symbolic Gr\"obner bases, we simply attempt to realize various $\mbf{K}(\ta)$ with randomly generated $\ta$. 
If successful, $\Scal$ is highly likely to be generically realizing; if not, it is highly likely to be not generically realizing.

\begin{algorithm}[h]
\textbf{Input:} $\Scal$ \\
\textbf{Output:} \tbf{True} or \tbf{False}
\caption{Determine if support $\Scal$ is generically realizing}
\label{alg:generically_realizing}
\begin{algorithmic}[1]
    \If{$\Scal$ does not satisfy the consequent of Theorem~\ref{thm:io_connection}}
    \State \Return{\tbf{False}}
    \EndIf
    \If{$\Scal$ does not satisfy the  consequent of Theorem~\ref{thm:gen_realizing_vs_struct_ctrb}}
    \State \Return{\tbf{False}}
    \EndIf
    \State Out = IsAccessible$(\Scal)$.
    \If{Out == \tbf{Accessible}}
    \State \Return{\tbf{True}}
    \EndIf
    \For{$T_p\in\Pcal_p$}
    \State Compute $H\coloneq H_{\Scal,a,T_p}$ for symbolic $\mbf{K}(a)$.
    \State $G = \up{ComputeReducedGr\"obnerBasis}(H,grevlex)$
    \If{$G\neq\{1\}$}
    \State \Return{\tbf{True}}
    \EndIf
    \EndFor
    \State \Return{\tbf{False}}
    
\end{algorithmic}
\end{algorithm}

\section{Numerical Examples} \label{sec:simulations}

We now present numerical examples that (1) realize a given controller on a given support, and (2) analyze whether a given support is generically realizing. 
All code required to reproduce our results are available on GitHub \footnote{https://github.com/YaozDu/Sparse-LTI-system-realization-via-solving-multivariate-polynomial-equations}.

\subsection{Realizing a controller on a given support}

\begin{figure}[h]
    \centering
    \includegraphics[width=0.6\linewidth]{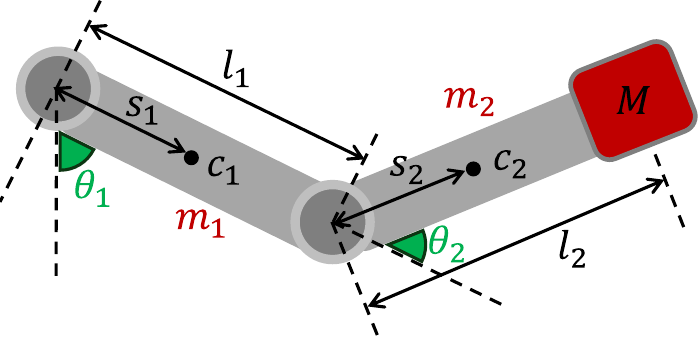}
    \caption{Two-joint robot arm model.}
    \label{fig:jointDynamic}
\end{figure}

We first construct a controller for a two-joint robot arm shown in Figure~\ref{fig:jointDynamic} with parameters given in Table~\ref{tab:joint_parameters}.
Here, $\theta_i$ represent joint angles; variable torque $\tau_1$ is applied to the first joint (this is the control input), and fixed torque $\tau_2 = 2.5$.

\begin{table}[h]
    \centering
    \begin{tabular}{|c|c|c|c|c|c|c|c|c|c|}
        \hline $I_1$ & 0.025 & $I_2$ & 0.045 & $m_1$ & 1.4 & $m_2$ & 1 & $M$ & 0.3\\
        \hline $l_1$ & 0.3 & $l_2$ & 0.33 & $s_1$ & 0.11 & $s_2$ & 0.16 & g & 9.8 \\
        \hline
    \end{tabular}
    \caption{Parameters of the joint dynamics model. $M$ is the mass of the payload [kg], $m_i$ is the mass of link $i$ [kg], $l_i$ is the length of link $i$ [m], $s_i$ is the distance between joint center and link $i$'s center of mass $c_i$ in [m], $I_i$ is link $i$'s moment of inertia [kg $\cdot \text{ m}^2$], and $g$ is gravitational acceleration [m/$\text{s}^2$].
    }
    \label{tab:joint_parameters}
\end{table}

Linearizing about fixed point $\theta_1 = \pi/4$, $\theta_2 = \pi/3$, and $\tau_1 = 6.2$ yields the following equations of motion:
\begin{equation}
\begin{split}
    \dot{x} &= Fx+Gu, \quad y = Hx, \\
    F &= \begin{bmatrix}
        0     &    0  &  1    &     0\\
        0     &    0     &    0   & 1\\
        -27.8  & -1.7   &     0    &     0\\
        44.7  &  8.7    &     0   &      0
    \end{bmatrix},
    G = \begin{bmatrix}
        0 \\0\\ 6.9\\-9.5
    \end{bmatrix} \\
    H &= \begin{bmatrix}
        0 & 1 & 0 & 0
    \end{bmatrix}
\end{split}
\end{equation}
where $x$ denotes the states (angles and angular velocities), and $u$ denotes input ($\tau_1$).
We design a stabilizing linear quadratic Gaussian (LQG) feedback controller for this plant: 
\begin{equation}
    \label{equ:Joint_Model_LQG_ctrller}
    \begin{aligned}
    \dot{z} &= (F+GK-LH)z+Ly,\quad u = Kz  \\
        K &= \begin{bmatrix}
            11.8  &  8.0 &   3.9   & 3.7
        \end{bmatrix}\\
        L &= \begin{bmatrix}
            -14.8  & 25.2 & -13.6 &  28.3
        \end{bmatrix}^\top    
    \end{aligned}
\end{equation}

\begin{figure}[h]
    \centering
    \begin{subfigure}[b]{0.35\columnwidth}
        \centering
        \includegraphics[width=\columnwidth]{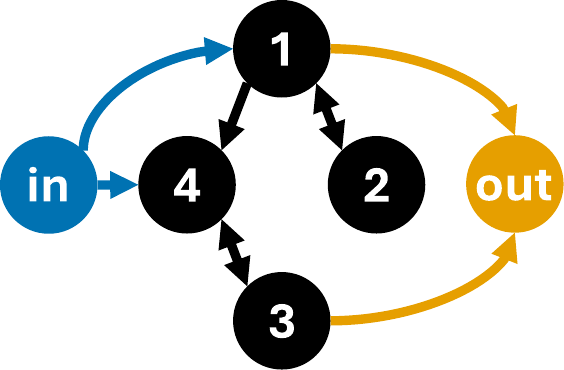}
        \caption{}
        \label{fig:1a}
    \end{subfigure}
    \hspace{2em}%\hfill
    \begin{subfigure}[b]{0.35\columnwidth}
        \centering
        \includegraphics[width=\columnwidth]{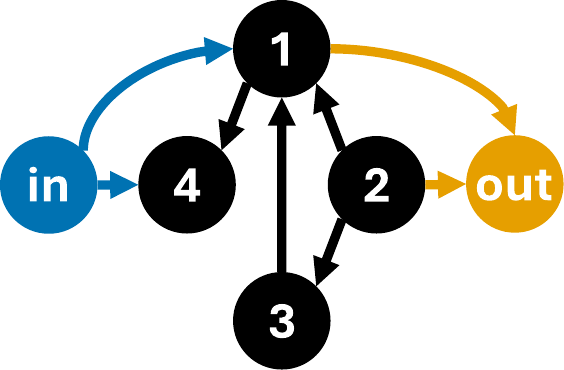}
        \caption{}
        \label{fig:1b}
    \end{subfigure}
    \caption{Controller connectivity patterns from numerical examples. Figure~\ref{fig:1a} corresponds to $\Scal_3$ in \eqref{equ:supp1}; Figure~\ref{fig:1b} corresponds to $\Scal_4$ in \eqref{equ:supp2}. For support $\Scal_i = (\Acal_i,\Bcal_i,\Ccal_i,0)$, $\Acal_i$ is the adjacency matrix for the internal nodes; $\Bcal_i$ and $\Ccal_i$ represent connections from input $y$ to the nodes and connections from nodes to output $u$, respectively.}
    \label{fig:main}
\end{figure}

Let $\mbf{K}(\ta)$ denote this controller. 
We first determine whether $\mbf{K}(\ta)$ is real-realizable on support $\Scal_3$ \eqref{equ:supp1}; this corresponds to Problem~\ref{pb:real_realization}.
\begin{equation}
\begin{aligned}
    \Acal_3 &= \begin{bmatrix}
        1   &  1    & 0   &  1\\
     1   &  1  &  0   &  0\\
     0   & 0   &  1  &   1\\
     0   &  0  &   1  &   1
    \end{bmatrix},\quad
    \tblue{\Bcal_3 = \begin{bmatrix}    1 \\ \tblue{0} \\ \tblue{0} \\ \tblue{1}  \end{bmatrix}} \\
    \torange{\Ccal_3} &\torange{= \begin{bmatrix}   1 & 0 & \torange{1} & \torange{0} \end{bmatrix}},\quad
    \Dcal = 0
\end{aligned}
\label{equ:supp1}
\end{equation}
This support is also shown in Figure~\ref{fig:1a}.

We compute $H_{\Scal_3,\ta,I_4}$, which contains 11 polynomials of 12 variables. 
Applying Algorithm~\ref{alg:find_realizations} with \tbf{complex} being false, we obtain the following realization:
\begin{equation}
    \begin{aligned}
        A_3 &= \begin{bmatrix}
            -17.4 &  -0.5  &  0 & -531651.7\\
            -2.5  &-17.4  & 0 &  0\\
            0  &  0 &  -17.2 & -457.3\\
            0  &  0  & 0.7  & 18.4
        \end{bmatrix} \\
    B_3 &= \begin{bmatrix}
        -37.6 & 0 & 0 & -0.3
    \end{bmatrix}^\top \\
    C_3 &= \begin{bmatrix}
        -2.1 & 0 & 2434.1 & 0
    \end{bmatrix}
    % I kept 2 digits after "."
    \label{equ:implementation_1}
    \end{aligned}
\end{equation}
This is a sparse realization of the dense controller \eqref{equ:Joint_Model_LQG_ctrller}.

We next determine whether $\mbf{K}(\ta)$ is realizable on support $\Scal_4$ \eqref{equ:supp2}; this corresponds to Problem~\ref{pb:complex_realization}.
\begin{equation}
    \begin{aligned}
        \Acal_4 &= \begin{bmatrix}
            1 & 0 & 0 & 1 \\ 1 & 1 & 1 & 0 \\ 1 & 0 & 1 & 0 \\ 0 & 0 & 0 & 1
        \end{bmatrix}, \quad
        \tblue{\Bcal_4 = \begin{bmatrix}    1 \\ 0 \\ 0 \\ 1\end{bmatrix}} \\
        \torange{\Ccal_4} &\torange{= \begin{bmatrix}     \torange{1} & \torange{1} & \torange{0} & \torange{0}   \end{bmatrix}},\quad
        \Dcal_4 = 0
        \label{equ:supp2}
    \end{aligned}
\end{equation}
This support is also shown in Figure~\ref{fig:1b}.
We compute $H_{\Scal_4,\ta,I_4}$, which contains 12 polynomials of 12 variables.
Applying Algorithm~\ref{alg:find_realizations} with \tbf{complex} being true, we obtain:
\begin{equation}
    \begin{aligned}
        A_4 &= \begin{bmatrix}
        3.3    &     0   &      0 &-337.8\\
    4.0 & -18.5 & -146.9  &       0\\
    0.3    &     0  & -16.2   &      0 \\
         0     &    0    &     0  & -2.2
        \end{bmatrix}, \\
        B_4 &= \begin{bmatrix}
            -712.9 & 0 & 0 & 0.9
        \end{bmatrix}^\top \\
        C_4 &= \begin{bmatrix}
            -0.1 & 1.1  & 0 & 0
        \end{bmatrix}
    \end{aligned}
    \label{equ:implementation_2}
\end{equation}

This is another sparse realization of the dense controller \eqref{equ:Joint_Model_LQG_ctrller}. Notice that we obtained a real-valued solution despite not being guaranteed to do so. We then consider a support $\Scal_5$ which is identical to $\Scal_3$ except with $\left[\Acal_5\right]_{41} = 1$, $\left[\Acal_5\right]_{12}=0$. 
We determine whether $\mbf{K}(\ta)$ is realizable on $\Scal_5$ by applying Algorithm~\ref{alg:find_realizations}. 
The algorithm returns ``Infeasible"; no complex or real-valued realizations exist.

\subsection{Determine if a support is generically realizing}

\begin{figure} 
 \centering
    \includegraphics[width=0.6\linewidth]{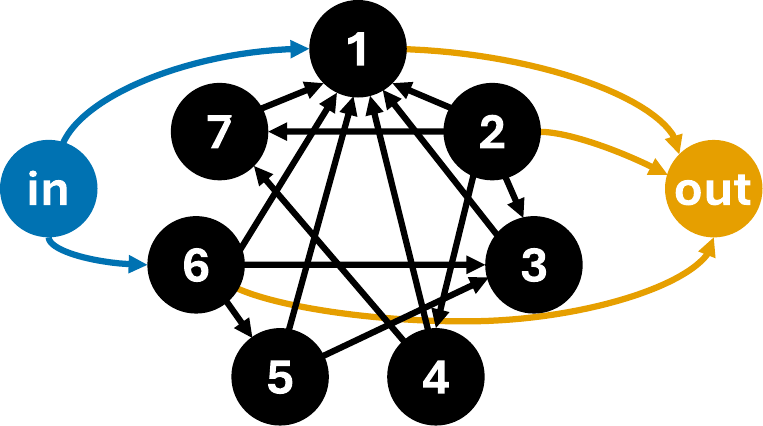}
    \caption{Connectivity pattern for $\Scal_6$.} \label{fig:support4}
\end{figure}

We now apply Algorithm~\ref{alg:generically_realizing} to determine whether supports $\Scal_3$, $\Scal_4$, $\Scal_5$ from the previous subsection are generically realizing.
We also analyze $\Scal_6$, depicted in Figure~\ref{fig:support4}. 
The results are listed in Table~\ref{tab:generic_realizability}.

\begin{table}[h]
    \centering
    \begin{tabular}{|c|c|c|}
        \hline support & accessible & generically realizing  \\
        \hline $\Scal_3$ & No & Yes \\
        \hline $\Scal_4$ & No & Yes \\
        \hline $\Scal_5$ & No & No \\
        \hline $\Scal_6$ & Yes & Yes \\
        \hline
    \end{tabular}
    \caption{Realizability of different supports}
    \label{tab:generic_realizability}
\end{table}

We find that $\Scal_3$ and $\Scal_4$ are not accessible; this requires us to compute symbolic Gr\"obner bases, which shows that they are generically realizing.
$\Scal_5$ is not generically realizing as it is not structurally controllable.
These results, in combination with the previous subsection's results, are reflective of Corollary~\ref{cor:not_gen_realizing} and the remarks following it.
That is, for an arbitrary controller, both the generically realizing supports were able to realize it, while the non-generically realizing support was not.
$\Scal_6$ is accessible and thus generically realizing; we do not need to compute symbolic Gr\"obner bases for it.
These results are also consistent with our finding that accessibility is sufficient but not necessary for generically realizing. 

\section{Conclusions and future work}
\label{sec:conclusions}

We first studied the problem of realizing a given controller on a support; we developed algorithms to provide complex- and real-valued realizations.
We then analyzed the set of controllers which can be realized on a given support. 
We showed that a support is either generically realizing (can realize nearly all controllers) or not (can realize nearly no controllers).
These analyses are achieved by converting the controller realization problem into a system of polynomial equations.
While comprehensive, our algorithms do not scale to large systems; this is the focus of future work.
We also plan to apply these algorithms to analyze neural circuit connectivity patterns (i.e., supports) from sensorimotor animal data.
%For future work, we plan to develop other efficient approaches for a restricted class of larger systems to determine on controller realizability and support's realization capacity, which provides numerical examples of larger systems.
%Potential future work include applying the framework for animal's movement regulation, which requires identifying supports from the sensorimotor system and identifying controller dynamics. We also plan to 

\bibliographystyle{IEEEtran}
\bibliography{References}

\begin{IEEEbiography}[{\includegraphics[width=1in,height=1.25in,clip,keepaspectratio]{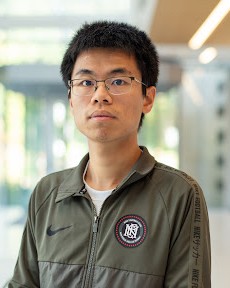}}]{Yaozhi Du}
    received the M.Sc. degree in Electrical and Computer Engineering from the University of Michigan, Ann Arbor, USA, in 2024, where he is currently working toward the Ph.D. degree. His research interests include structured controller design and controller realization.
\end{IEEEbiography}

\begin{IEEEbiography}[{\includegraphics[width=1in,height=1.25in,clip,keepaspectratio]{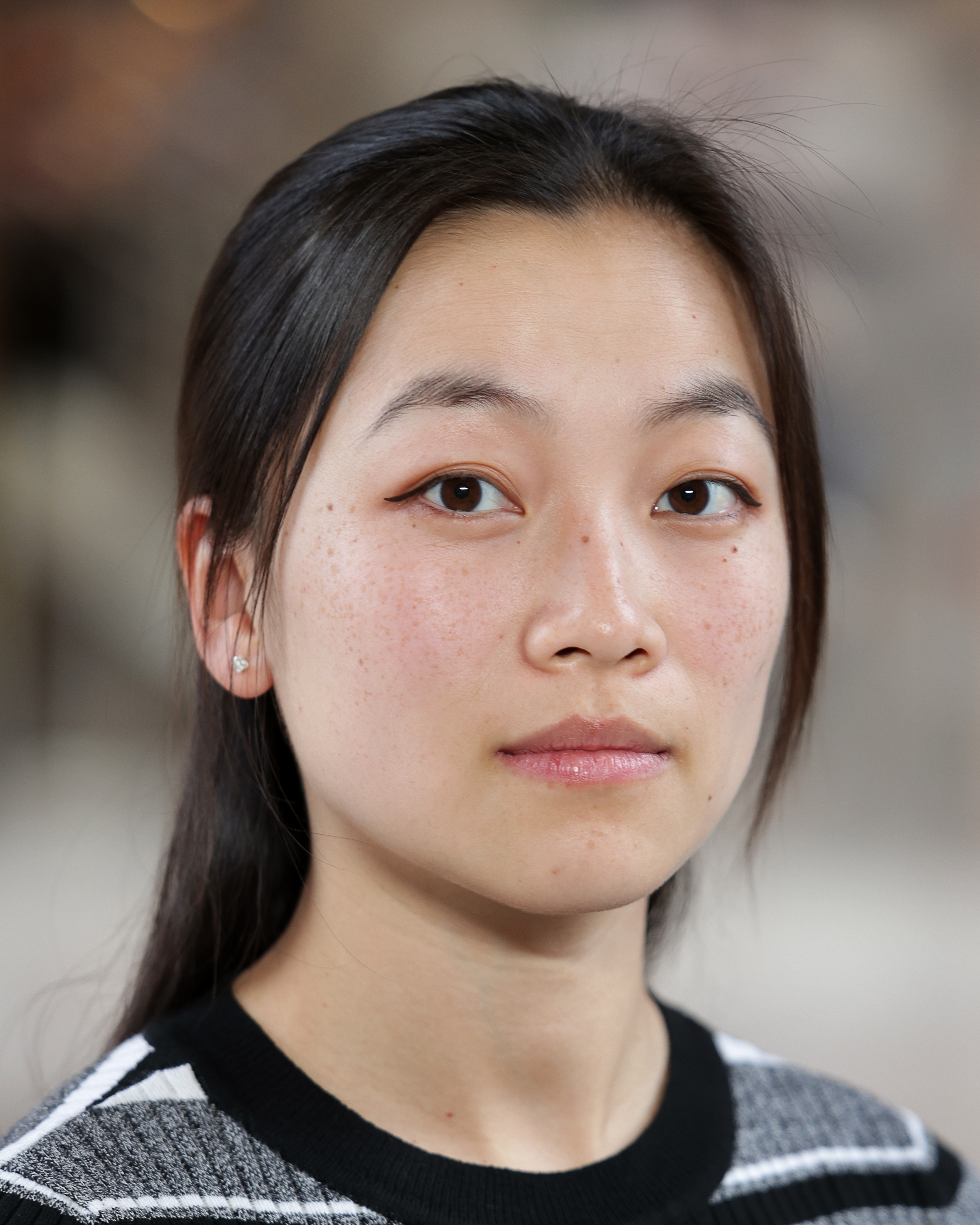}}]{Jing Shuang (Lisa) Li}
    received the Ph.D. in Control and Dynamical Systems from the California Institute of Technology, Pasadena, USA, in 2023; she is currently an Assistant Professor at the University of Michigan, Ann Arbor, USA. Her research interests include distributed control, resource-constrained control, computation via dynamics, and large-scale systems, with applications in both engineering and neuroscience.
\end{IEEEbiography}

\end{document}